\documentclass[12pt]{article}
\usepackage{amsmath,amssymb}

\usepackage{graphicx}
\usepackage{wrapfig}

\usepackage{hyperref}
\usepackage{cite}

\def\beq{\begin{eqnarray}}
\def\eeq{\end{eqnarray}}

\def\o{\over}

\newcommand{\spt}{s}         
\newcommand{\apt}{\sigma}      

\newcommand{\CMdim}[1]    {{\mathfrak{M}^{#1}}} 

\newcommand{\pBt}{{\Delta_t}}
\newcommand{\pBx}{{\Delta_x}}

\newcommand{\Nl}{\mathrm{N}}      

\newcommand{\Tr}{\,\mathrm{Tr}\,}            

\newcommand{\za}{{\alpha}}   
\newcommand{\zb}{{\beta}}    
\newcommand{\zm}{{\mu}}      
\newcommand{\zn}{{\nu}}      

 \newcommand{\GammaF}[1] {{\ensuremath{\mathchoice%
      {\,\mathrm{\Gamma}{\textstyle\left(#1\right)}}
      {\,\mathrm{\Gamma}{\textstyle\big(#1\big)}}
      {\,\mathrm{\Gamma}{\scriptstyle(#1)}}
      {\,\mathrm{\Gamma}{\scriptscriptstyle\left(#1\right)}} }}}

\newcommand{\fract}[2]{{\textstyle\frac{#1}{#2}}}

\newcommand{\be}{\begin{equation}}
\newcommand{\ee}{\end{equation}}
\newcommand{\bea}{\begin{eqnarray}}
\newcommand{\eea}{\end{eqnarray}}
\newcommand{\bg}{\begin{gather}}
\newcommand{\eg}{\end{gather}}
\newcommand{\bseq}{\begin{subequations}}
\newcommand{\eseq}{\end{subequations}}

\renewcommand{\tanh}{\mathop{\rm th}\nolimits}

\renewcommand{\ln}{\mathop{\rm ln}\nolimits}

\def\be{\begin{eqnarray}}
\def\ee{\end{eqnarray}}
\def\lb{\label}

\begin{document}

\title{\textbf{Gravitational effective action and entanglement entropy in UV modified theories with and  without Lorentz symmetry}}

\vspace{1cm}
\author{ \textbf{Dmitry Nesterov$^{\star}$ and
 Sergey N. Solodukhin$^\sharp$ }} 

\date{}
\maketitle

\begin{center}
  \hspace{-0mm}
  \emph{$^{\star}$ $^{\sharp}$ Laboratoire de Math\'ematiques et Physique Th\'eorique }\\
  \emph{Universit\'e Fran\c cois-Rabelais Tours,\\ F\'ed\'eration Denis Poisson - CNRS, }\\
  \emph{Parc de Grandmont, 37200 Tours, France} \\
  \emph{ and }\\
  \emph{$^{\star}$ Theory Department, Lebedev Physical Institute,
      \\ Leninsky Prospect 53, Moscow, Russia, 119991.}
\end{center}

{\vspace{-11cm}
\begin{flushright}
\end{flushright}
\vspace{11cm}
}



\begin{abstract}
\noindent { We calculate parameters in the low energy gravitational effective action  and the entanglement entropy in a wide class of theories characterized by improved ultraviolet (UV) behavior. These include i) local and non-local Lorentz invariant theories in which inverse propagator is modified by higher-derivative terms and ii) theories described by non-Lorentz invariant Lifshitz type field operators. We demonstrate that the induced cosmological constant, gravitational couplings and the  entropy are sensitive to the way the theory is modified in UV. For non-Lorentz invariant  theories the induced gravitational effective action is of the Horava-Lifshitz type.  We show that under certain conditions imposed on the dimension of the Lifshitz operator the couplings  of the  extrinsic curvature terms in the effective action are UV finite. Throughout the paper we systematically exploit the heat kernel method appropriately generalized for the class of theories under consideration.}
\end{abstract}

\vskip 2 cm
\noindent
$^{\star}$ e-mail: nesterov@lpi.ru, Dmitry.Nesterov@lmpt.univ-tours.fr \\
$^{\sharp}$ e-mail: Sergey.Solodukhin@lmpt.univ-tours.fr

\newpage
    \tableofcontents
\pagebreak

\newpage

\section{ Introduction}
\setcounter{equation}0

It is believed that the transplanckian physics manifests itself in certain modifications of the existing field models that makes the renormalizability of the quantum models more obvious if not  automatic. String theory which at the present stage is the only consistent quantum theory which unifies gravity with other interactions effectively provides a mechanism for such a modification. Yet, the concrete details of this mechanism  remain to be understood\footnote{For gravitational field, however, we note that there exists an alternative point of view recently phrased in \cite{Dvali:2010bf}  that pure Einstein Gravity is already UV complete in the way that transplanckian degrees of freedom cannot exist.}.
One possible way of how this mechanism may work is via modifications in the field propagator so that, as  function of momentum, it decays faster  than the standard $1/p^2$. The loop diagrams then will be regularized  and one even may hope to get a theory free of UV divergences.  In this paper we study the consequences of this improved UV behavior of field propagator for quantities which appear in the effective gravitational action and determine the gravitational interaction as we know it from the every day experience and astronomical observations.
These quantities are cosmological constant and Newton's constant. More precisely we study how the UV divergences of these quantities are modified by the modifications of the propagator. In particular, we investigate whether those divergences can be completely removed by appropriate choice of the modified propagator.

For simplicity we consider the case of a scalar field. Generalization to other fields is straightforward.
We consider two wide classes of theories. The first class includes local and non-local Lorentz invariant theories. The inverse propagator in this case can be represented in the following general form
 \begin{equation} \lb{i1}
  G^{-1}(p)=F(p^2) \,,
 \end{equation}
where $F(p^2)$ is some (to be specified) function of momentum\footnote{Throughout this paper we work in the Euclidean signature so that $p^2\geq 0$.} $p^2=p^\mu p_\mu$. As a typical example of theories of this type
we consider a theory with higher derivatives. The inverse propagator then is a polynomial of degree $k$ of $p^2$,
 \begin{equation} \lb{i2}
  F(p^2)=p^2+{p^{2k}\over m^{2(k-1)}}~.
 \end{equation}
The theory may include infinite number of derivatives, for example
 \begin{equation} \lb{i3}
  F(p^2)=p^2+m^2e^{p^2\over \Lambda}~.
 \end{equation}
The theories with higher derivatives are usually considered as not well defined due to presence of ghosts.  In this paper we do not address this issue. We note however that a number of approaches have been suggested recently to properly treat higher derivative theories  \cite{HH}, \cite{Antoniadis:2006pc}, \cite{Bender}, \cite{Smilga:2005gb}.

The second class of theories are theories with broken Lorentz symmetry. These are Lifshitz type theories \cite{Lifshitz}, \cite{Horava}. The inverse propagator in this case is
 \begin{equation} \lb{i4}
  G^{-1}(\omega, {\bf p})=\omega^2+F({\bf p}^2)~,
 \end{equation}
where the momentum now splits on time-like and space-like components, $p^\mu=(\omega,{\bf p}^i)$.
A typical example of function $F$ in this case is
 \begin{equation} \lb{i5}
  F({\bf p}^2)={\bf p^2}+{{\bf p}^{2k}\o m^{2(k-1)}}~.
 \end{equation}
The second term in (\ref{i5}) breaks the Lorentz symmetry for large values of space-like momentum $\bf p$ while the symmetry is restored in the regime of small values of $\bf p$ (compared to the mass scale $m$).

The other quantity which we study in this paper and which is anticipated to be sensitive to the UV modifications of the theory is entanglement entropy. Entanglement entropy of a field system is defined by tracing over modes  that reside  inside a surface $\Sigma$. If the total system is characterized by a pure quantum state the subsystem inside the surface is described by a density matrix $\rho$ with the von Neumann entropy $S=-\Tr \rho\ln\rho$ known as {\it entanglement entropy} (see \cite{BS}, \cite{Srednicki:1993im} recent reviews \cite{ee}, \cite{Casini:2009sr} and references therein). This entropy is non-vanishing provided there are short-distance correlations in the system. The presence of such correlations has two consequences: a) entanglement entropy is determined by geometry of the surface $\Sigma$, to leading order by the area of $\Sigma$ and in higher orders by intrinsic and extrinsic geometry of $\Sigma$; b) the entropy is UV divergent, in $d$ spacetime dimensions in a theory with propagator $1/p^2$ one has that
 \begin{equation} \label{S}
  S\sim N{Area ( {\Sigma}) \o \epsilon^{d-2}}.
 \end{equation}
The exact coefficient of proportionality in (\ref{S}) depends on the regularization scheme employed, $N$ is the number of species. Roughly speaking, the short-distance correlation $1/\sigma^{d-2}$ between 2 points separated by interval $\sigma$ in $d$ dimensions is reflected in the $1/\epsilon^{d-2}$ divergence of the entropy. Entanglement entropy thus provides us with information about the structure of the quantum field theory in UV regime \cite{Dvali:2008jb}. By improving the UV behavior of the theory one would get a milder divergence of the entropy. It is believed that in string theory the entropy should be completely finite \cite{Susskind:1994sm}. One of the goals of this paper is to test this statement on the theories with modified UV behavior, (\ref{i1}) and (\ref{i4}).

Although we are primarily interested in theories with  $F(p^2)$ growing faster than $p^2$ our results are general and applicable to other cases as well. It should be  noted that theories in which propagator is a non-trivial function of momentum are not rare. Let us remind the reader some of theories of this type:

i) 4-dimensional brane in spacetime with extra dimensions, for instance in a theory with one extra dimension the induced propagator along the brane is described by function \cite{Barvinsky:2003jf}
 \begin{equation}
  F(p^2)={p\over L}\tanh(Lp)~,
 \lb{i6}
 \end{equation}
where $L$ is the size of extra dimension.

ii) DGP model \cite{DGP}, the inverse propagator induced on the brane then can be modeled by the function \cite{Barvinsky:2006pg}
 \begin{equation}
  F(p^2)=p^2+m\sqrt{p^2}~.
 \lb{i7}
 \end{equation}

iii) non-commutative field theories,
 \begin{equation}
  F(p^2)=p^2+{1\over \theta^2p^2}~.
 \lb{i8}
 \end{equation}

It should be noted that in the last two examples the theory is modified in infra-red (IR) rather than in the UV regime.

This paper is organized as follows. We first consider the Lorentz invariant theories. In Section~2  we compute the vacuum energy in a theory of the type (\ref{i1}). This computation is done in flat spacetime using the momentum representation of the corresponding heat kernel function. Covariant calculations in curved spacetime confirm this result, which should be interpreted as induced cosmological constant. Entanglement entropy in Lorentz invariant theories is computed in Section~3. This computation is  done in flat space-time as well. The curved spacetime is considered in Section~4 where we compute the induced  Newton's constant. In Section~5 we present general results on the structure of the heat kernel for an operator which is function of the Laplace operator. Non-Lorentz invariant theories are considered in Section~6, where we first present the general structure of the low-energy gravitational effective action and use the scaling arguments to analyze the structure of UV divergences. Then we present a calculation of entanglement entropy in this class of theories. We finish Section~6 by giving a precise form for the dependence of couplings in the effective action on the UV cutoff. Some technical details are collected in Appendices.

\section{Vacuum Energy/Cosmological Constant}

\setcounter{equation}0 As a warm-up we calculate the vacuum energy density in flat spacetime in a field theory with  field operator
 \begin{equation} \lb{1}
  {\cal D}=F(\Box)~,
 \end{equation}
which is  an arbitrary function of the Laplace operator $\Box=-\partial_\mu\partial^\mu$. In this paper we systematically use the method of the heat kernel in order to calculate the effective action. Consider a quantum bosonic field  described by a field operator $\cal D$ so that the partition function is $Z=\det^{-1/2}{\cal D}$. Then the effective action defined as
 \begin{equation}
  W=-{1\o 2}\int_{\epsilon^2}^\infty {ds\o s}\Tr K (s)\lb{W}\,,
 \end{equation}
where parameter $\epsilon$ is an UV cutoff, is expressed in terms of the trace of the heat kernel $K(s,X,X')=<X|e^{-s{\cal D}}|X'>$. The latter is defined as a solution  to the heat  equation
 \begin{equation}\label{K}
  \left\{
  \begin{array}{l}
    (\partial_s+{\cal D}) \,K(s,X,X')=0 \,, \\
    K(s{=}0,X,X')=\delta(X,X') \,.
  \end{array}
    \right.
 \end{equation}

In flat spacetime one can use the Fourier transform  in order to solve the heat equation. In $d$ spacetime dimensions one has
 \begin{equation}
  K(s,X,X')={1\over (2\pi)^d}\int d^dp \,e^{ip_\mu(X^\mu{-}X'^\mu)}~e^{-sF(p^2)}\,.
 \lb{2}
 \end{equation}
In this representation the trace of the heat kernel $\Tr K(s)=\int d^dX\, K(s,X,X'{=}X)$ reduces to a single integral
 \begin{equation}
  \Tr K(s)= \frac1{(4\pi)^{d/2}} V  P_{d}(s)\,,
 \lb{3}
 \end{equation}
where
$V=\int d^dX$ is the volume of the spacetime and we introduced factors
 \begin{equation}
  P_{n}(s)=\frac2{\GammaF{\frac{n}2}} \int_0^\infty dp\, p^{n-1}\, e^{-s F(p^2)}\,,
 \label{4}
 \end{equation}
which in the case of Laplace operator ${\cal D}=\Box$ and hence $F(p^2)=p^2$ give just $s^{-n/2}$.

The effective action
 \begin{equation}
  W=-\lambda(\epsilon) \, V\,,
 \end{equation}
in this case represents the vacuum energy with the density
 \begin{equation}
  \lambda(\epsilon)
  = \frac12 \frac1{(4\pi)^{d/2}} \int_{\epsilon^2}^\infty \frac{ds}{s} \,P_{d}(s)~,
 \lb{5}
 \end{equation}
where $\epsilon$ is an UV cutoff which is introduced to regularize the UV divergences. The small $s$ behavior of $P_n(s)$ is determined by the behavior of function $F(p^2)$ for large values of $p$. Clearly, the functions $P_n(s)$ are always divergent in the limit of small $s$ no matter how fast
function $F(p^2)$ is growing with $p$. Respectively the vacuum
energy $\lambda(\epsilon)$ (\ref{5}) is  UV divergent for any function
$F(p^2)$. The divergence however can be milder depending on the
function $F(p^2)$. In curved spacetime the quantity $\lambda(\epsilon)$ (\ref{5})  should be identified with the cosmological constant.
\\

As an elementary particular case consider the standard field operator $F(\Box)=\Box$  in $d=4$ for which one gets
 \begin{equation} \lb{*}
  P_4(s)=\frac{1}{s^2}\;, \quad \lambda(\epsilon)=\frac{1}{64\pi^2} \frac{1}{\epsilon^{4}}\,.
 \end{equation}
The vacuum energy  in this case  is determined only by the UV cutoff $\epsilon$. This is standard UV divergence of cosmological constant in four spacetime dimensions.

The UV modification  of the field operator however alters the structure of the vacuum energy.
Consider two demonstrative examples.
\\

\noindent {\it Example 1: The function $F(p^2)$ is polynomial in
$p^2$.} \\
For simplicity we take $d=4$ and
 \begin{equation}
   F(p^2)=p^2+p^4/m^2~,
  \lb{6}
  \end{equation}
where $m$ is a new scale that characterizes the UV completion of the theory. In this case we have that
 \begin{equation} \lb{7}
  P_4(s)=\frac{m^2}{2s}
  \left(1+ \sqrt{\pi} \frac {m \sqrt{s}}{2} e^{{s m^2\o 4}} \big(\Phi({m\sqrt{s}\o 2})-1\big)\right)\,,
 \end{equation}
where $\Phi(x)$ is error function $\frac2{\sqrt{\pi}} \int_0^x dt\,e^{-t^2}$ \cite{GradRyzh}.

In the limit of small $s$ such that $m^2s\ll 1$ the function $P_4(s)$ behaves as
 \begin{equation} \lb{8}
  P_4(s)= \frac{m^4}{8}\left(\frac{4}{sm^2}-\sqrt{4\pi\o sm^2}+2+O(m\sqrt{s})\right)\,.
 \end{equation}
Respectively, the vacuum energy reads
  \begin{equation} \lb{9}
  \lambda(\epsilon)={1\o 16(2\pi)^3}\left({m^2\over \epsilon^2}
     -{\sqrt{\pi}}{m^3\o \epsilon}-{m^4} \ln\big({m\epsilon\o 2}\big)\right)~,
 \end{equation}
where we omitted terms which are finite in the limit $\epsilon\rightarrow 0$. The leading divergence is now quadratic. There are also  linear and logarithmically divergent terms in the vacuum energy.

As could be anticipated, in the regime $sm^2\gg 1$ and $\epsilon m \gg 1$ the function $P_4(s)$ and the vacuum energy for the field operator (\ref{6}) reproduce the result (\ref{*}) for the standard field operator.

In the limit of small $s$ in \emph{arbitrary} dimension $d$ the integral $P_{d}(s)$ is dominated by large values of $p$. Suppose that for large $p$ the function
  \[F(p^2)\sim p^{2k}/m^{2(k-1)}\,, \]
or, in terms of the field operator, ${\cal D}\sim \Box^k/m^{2(k-1)}$. Then the heat kernel $e^{-s{\cal D}}$
is invariant under the following rescalings
 \begin{equation} \lb{99}
  X^\mu\rightarrow \beta^{1/k} X^\mu\,,\quad
  s\rightarrow \beta^2 s \,.
 \end{equation}
This symmetry tells us that  in $d$ spacetime dimensions
 \begin{eqnarray}
  &&\Tr K(s) \sim {V \o s^{d\o 2k}}\,, \qquad s\rightarrow 0\,,
 \nonumber\\
  &&W = -\lambda(\epsilon)\, V\,, \quad \lambda(\epsilon)\sim { m^{d(k-1)\o k}\o \epsilon^{d\o k}}\,.
 \lb{999}
 \end{eqnarray}
\\

\noindent{\it Example 2: $F(p^2)$ is  exponentially growing function of $p^2$.}
\\
We take
 \begin{equation} \lb{10}
  F(p^2)=m^2(e^{p^2/\Lambda^2}-1)\,.
 \end{equation}
It is characterized by two parameters, $m$ and $\Lambda$. Our choice for the function $F(p^2)$ is such that the
pole of the propagator $F^{-1}(p^2)$ is at $p^2=0$, so that in the infra-red regime we have a massless field equation.

The integral $P_4(s)$ for the function (\ref{10}) is calculated explicitly and is expressed in terms of the hypergeometric function. In the limit of small $s$ such that $s m^2\ll 1$ we have
 \begin{equation} \lb{11}
  P_4(s)=\Lambda^4\left({\pi^2\o 12}+{\gamma^2\o 2}+\gamma \ln sm^2+{1\o 2}(\ln sm^2)^2+O(s\ln^2s)\right)\,,
 \end{equation}
where $\gamma$ is the Euler's constant, so that the vacuum energy
 \begin{equation} \lb{12}
  \lambda(\epsilon)=\Lambda^4 \left({2\o 3}(\ln{1\o \epsilon m})^3-\gamma(\ln{1\o \epsilon m})^2+({\pi^2\o 12}+{\gamma^2\o 2})\ln{1\o \epsilon m}\right)
 \end{equation}
diverges logarithmically in $\epsilon$. For $m$ of order of the cutoff the cosmological constant is determined by scale $\Lambda$ which can be considerably smaller than the cutoff. Thus the problem of cosmological constant should be reconsidered in the UV modified theories.

\bigskip
\section{Entanglement Entropy}
\setcounter{equation}0
\subsection{ The replica method} Before proceeding we remind a
technical method very useful for calculation of entanglement
entropy in a field theory described by the Laplace operator.
This method is known as {\it the replica trick}, see
ref.\cite{Callan:1994py}.
Entanglement entropy is associated with a surface $\Sigma$ and is
defined by tracing the modes of quantum fields residing inside the
surface. One considers a quantum field $\psi(X)$ on
$d$-dimensional spacetime and  chooses the Cartesian coordinates
$X^\mu=(\tau,x, z^i,\, i=1,..,d{-}2)$ where $\tau$ is Euclidean time,
such that the surface $\Sigma$ is defined by condition $x=0$ and
$(z^i,\, i=1,..,d{-}2)$ are the coordinates on $\Sigma$. In the
subspace $(\tau,x)$ it will be  convenient to choose the polar
coordinate system $\tau=r\sin(\phi)$ and $x=r\cos(\phi)$ where
angular coordinate $\phi$ changes in the limits $0\leq \phi <
2\pi$.

One first defines a vacuum state of the quantum field in question
by the path integral over a half of the total Euclidean spacetime
defined as $\tau\leq 0$ such that the quantum field takes the
fixed boundary condition $\psi(\tau=0,x,z)=\psi_0(x,z)$ on the
boundary of the half-space,
 \begin{equation}
  \Psi[\psi_0(x,z)]=\int\limits_{\psi(X)|_{\tau=0}=\psi_0(x,z)}
   {\cal D}\psi\; e^{-W[\psi]}~,
  \lb{W1}
  \end{equation}
where $W[\psi]$ is the action
of the field. The surface $\Sigma$ in our case is a plane and
direction of the coordinate $x$ is orthogonal to $\Sigma$. So that
the surface $\Sigma$ defined by condition $x=0,\; \tau=0$
naturally separates   the hypersurface $\tau=0$ on two parts:
$x<0$ and $x>0$. Respectively the boundary data $\psi(x,z)$ is
separated on $\psi_-(x,z)=\psi_0(x,z),\; x<0$ and
$\psi_+=\psi_0(x,z),x>0$. Tracing over $\psi_-(x,z)$ defines a reduced
density matrix
 \begin{equation}
  \rho(\psi_+^1,\psi_+^2)=\int {\cal D}\psi_-\Psi(\psi_+^1,\psi_-)\Psi(\psi_+^2,\psi_-)~,
 \lb{W2}
 \end{equation}
where the path integral goes over fields defined on the whole
Euclidean spacetime except a cut $(\tau=0,x>0)$. In the path
integral the field $\psi(X)$ takes a boundary value $\psi_+^2$
above the cut and $\psi_+^1$ below the cut. The trace of $n$-th
power of the density matrix (\ref{W2}) then is given by the
Euclidean path integral over fields defined on an $n$-sheeted
covering  of the cut spacetime. In the polar coordinates
$(r,\phi)$ the cut corresponds to values $\phi=2\pi k,$
$k=1,2,..,n$. When one passes through the cut from one sheet to
another the fields are glued analytically. Geometrically this
$n$-fold space is a flat cone $C_n$ with angle deficit $2\pi(1-n)$
at the surface $\Sigma$. Thus we have
 \begin{equation}
  \Tr\rho^n=Z[C_n]~,
 \lb{W3}
 \end{equation}
where $Z[C_n]$ is the Euclidean path integral over the $n$-fold cover of the Euclidean space, i.e. over the cone $C_n$.

Assuming that in (\ref{W3}) one can analytically continue to
non-integer values of $n$ one  observes that $-\Tr \hat{\rho} \ln
\hat{\rho}=-(\alpha\partial_\alpha-1)\ln\Tr
\rho^\alpha|_{\alpha=1}$, where $\hat{\rho}={\rho\o \Tr\rho}$ is
the renormalized matrix density. Introduce the effective action
$W[\alpha]=-\ln Z(\alpha)$, where $Z(\alpha)=Z[C_\alpha]$ is the
partition function of the field system in question on a Euclidean
space with conical singularity at the surface $\Sigma$. In the
polar coordinates $(r,\phi)$ the conical space $C_\alpha$  is
defined by making the coordinate $\phi$ periodic with the period
$2\pi\alpha$, where $(1-\alpha)$ is very small. Thus one has that
 \begin{equation}  \lb{SS}
  S=(\alpha\partial_\alpha-1)W(\alpha)|_{\alpha=1}\,.
 \end{equation}

The uniqueness of the analytic continuation of $\Tr \rho^n$ to
non-integer $n$ may seem not obvious, especially if the field
system in question is not relativistic so that there is no
isometry in the polar angle $\phi$ which would allow us without
any troubles to glue together pieces of the Euclidean space to
form a   path integral  over a conical space $C_\alpha$. However,
the arguments are given that the analytic continuation to
non-integer $n$ is in fact unique. These arguments in some form
 have already appeared in the literature \cite{CC}. Here we briefly reproduce
some version of these arguments since this issue becomes  important when we consider the
quantum fields with broken Lorentz symmetry.

Consider a renormalized density matrix $\hat{\rho}={\rho\o \Tr
\rho}$. The eigenvalues of $\hat{\rho}$ lie in the interval
$0<\lambda<1$. If this matrix was a  finite matrix we could use
the triangle inequality to show that
 \[ |\Tr \hat{\rho}^\alpha|<|(\Tr\hat{\rho})^\alpha|=1~
   \quad\text{if}\quad Re(\alpha)>1\;. \]
For infinite size matrices  the trace is usually infinite so that
a regularization is needed to  regularize it. Suppose $\epsilon$
is a such regularization parameter and $\Tr_\epsilon$ is the
regularized trace. Then
 \be
  |\Tr_\epsilon\hat{\rho}^\alpha|<1~\quad \text{if}\quad Re(\alpha)>1\;.
 \lb{W5}
 \ee
Thus $\Tr\hat{\rho}^\alpha$ is a bounded function on

the complex half-plane, $Re(\alpha)>1$. In fact, the path integral calculation
on a conical space (see next section) gives
 \begin{equation} \lb{W6}
  \Tr_\epsilon{\hat{\rho}}\sim e^{-{(\alpha^2-1)\o \alpha}S(\epsilon)} \,,
 \end{equation} where
$S(\epsilon)$ is the entanglement entropy. Clearly (\ref{W6})
satisfies (\ref{W5}). Now suppose we know $\Tr_\epsilon
\rho^\alpha|_{\alpha=n}=Z_0(n)$ for integer values of $\alpha=n,\;
n=1,2,3,..$. Then in the region $Re(\alpha)>1$ we can represent
$Z(\alpha)=\Tr_\epsilon\rho^\alpha$ in the form
 \begin{equation} \lb{W7}
  Z(\alpha)=Z_0(\alpha)+\sin(\pi\alpha)g(\alpha) \,,
 \end{equation}
where the function $g(\alpha)$ is  analytic (for
$Re(\alpha)>1$). Since by condition (\ref{W5}) the function
$Z(\alpha)$ is bounded we obtain that in order
to compensate the growth of the sinus in (\ref{W7}) for complex
values of $\alpha$ the function $g(\alpha)$ should satisfy
condition
 \begin{equation} \lb{W8}
  |g(\alpha=x+iy)|<e^{-\pi |y| }~.
 \end{equation}
By Carlson's theorem \cite{Carlson} an analytic function  bounded in the region
$Re(\alpha)>1$ and  which satisfies condition (\ref{W8})
vanishes identically. Thus we conclude that $g(\alpha)\equiv0$
and there is only one analytic continuation to
non-integer $n$ namely the one given by function $Z_0(\alpha)$.

In order to calculate the effective action $W(\alpha)$ we use the heat kernel method. In the context of the manifolds with conical singularities this method was earlier developed in great detail in \cite{coneQFT}, \cite{Fursaev:1994in}. In the Lorentz invariant case the heat kernel $K(s,\phi,\phi')$ (where we skip the coordinates other than angle $\phi$) on regular flat space depends on the difference $(\phi-\phi')$. The heat kernel $K_\alpha (s,\phi,\phi')$ on space with a conical singularity is then constructed from this quantity by applying the Sommerfeld formula
\cite{Sommerfeld}
  \begin{equation}
   K_\alpha(s,\phi,\phi')
    = K(s,\phi-\phi')  + { i \o 4\pi\alpha}\int_\Gamma \cot {w\o 2\alpha} K(s,\phi-\phi'+w)dw \,.
  \label{Sommerfeld}
  \end{equation}
The contour $\Gamma$
consists of two vertical lines, going from $(-\pi+ i \infty )$
to $(-\pi- i \infty )$ and from $(\pi- i \infty )$ to
$(\pi + i \infty )$ and intersecting the real axis between the
poles of the $\cot {w\o 2\alpha}$: $-2\pi\alpha$, $0$ and $0,$
$+2\pi\alpha$ respectively. For $\alpha=1$ the integrand in
(\ref{Sommerfeld}) is a $2\pi$-periodic function and the
contributions of these two vertical lines cancel each other. Thus,
for a small angle deficit the contribution of the integral in
(\ref{Sommerfeld}) is proportional to $(1-\alpha)$.

\subsection{The Lorentz invariant theories}
In this section we want to adopt the definition of entanglement
entropy to a  generic Lorentz invariant non-interacting theory
described by a field operator $\cal D$  \begin{equation} {\cal D}=F(\Box)
\lb{F}~, \end{equation} which can be an arbitrary function of the operator
$\Box=-\partial^\mu\partial_\mu$. We have to construct the
vacuum state of the theory, an analogue of the state (\ref{W1}).
Suppose that operator (\ref{F}) is a polynomial of degree $N$ of Laplace operator $\Box$.
First we note that one needs to specify $N$ functions on the boundary at $\tau=0$, $\psi_1(z,x)=\psi(\tau=0,z,x)$,
$\psi_2(z,x)=\partial_\tau\psi(\tau=0,z,x),.., \psi_N(z,x)=\partial^N_\tau\psi(\tau=0,z,x)$. This is the right set of data needed to uniquely define the eigenfunctions of operator $\cal D$ on a space with boundary at $\tau=0$.

Thus, we define the vacuum state by
 \be
  \Psi[\{\psi_a,\; a=1..N\}]=\int\limits_{\{\psi_a(z,x),\; a=1..N\}}D\psi(\tau,z,x) \; e^{-{1\o 2}\int \psi^T\hat{\cal D}\psi}~.
 \lb{L4}
 \ee
This prescription for the  state is consistent with the one introduced by Hawking and Hertog \cite{HH}.
The reduced density matrix is defined as before by tracing over all functions $\{\psi_a,\; a=1..N\}$
defined on the left half, $x<0$, of the hypersurface $\tau=0$. The
trace of $n$-th power of the density matrix is then given by the unrestricted
path integral over fields defined on the $n$-sheeted cover of the Euclidean space,
$Z[C_n]={\det}^{-1/2}\hat{ D}$.

In order to calculate the effective action $W[C_n]=-\ln Z[C_n]$
for partition function $Z[C_n]={\det}^{-1/2}F(\Box)$ we use the
heat kernel. As we have already seen, in flat Euclidean space the heat kernel for  operator
$F(\Box)$ is obtained by the Fourier transform
 \begin{equation}
   K(s,X,X')={1\o (2\pi )^d}\int d^dp\, e^{ i p_\mu (X^\mu-X'^\mu )}e^{-sF(p^2)}~.
 \lb{KF}
 \end{equation}
Putting $z^i=z'^i,\, i=1,..,d-2$  and choosing in the polar coordinate system $(r,\phi)$ that $\phi=\phi'+w$ we have that $p_\mu(X{-}X')^\mu=2pr\sin{w\o 2}\cos\theta$, where  $p^2=p^\mu p_\mu$ and $\theta$ is the angle between $d$-vectors $p^\mu$ and $(X^\mu{-}X'^\mu)$. Radial momentum $p$ and angle $\theta$ together with  $(d{-}2)$ other angles  form a spherical coordinate system in the space of momenta $p^\mu$. Thus one has for the integration measure $\int d^dp=\Omega_{d-2}\int_0^\infty dp\,p^{d-1}\int_0^\pi d\theta\,\sin^{d-2}\theta\,$\, where $\Omega_{d-2}={2\,\pi^{(d{-}1)/2}\o \Gamma((d{-}1)/2)}$ is the area of a unit radius sphere in $d{-}1$ dimensions. Performing the integration in (\ref{KF}) in this coordinate system we find
 \begin{equation}
  K(s,w,r)={\Omega_{d-2}\sqrt{\pi}\o (2\pi)^d} \frac{\GammaF{\frac{d{-}1}{2}}}{(r \sin{w\o 2})^{(d-2)/ 2}} \int_0^\infty\!\! dp \, p^{d\o 2}J_{d-2\o 2} \big(2rp\sin{\fract{w}{2}}\big)\,e^{-sF(p^2)} \,.
 \lb{Kw}
 \end{equation}
We used the fact that
 \[ \int_0^\pi \!\! d\theta\, \sin^{d{-}2}\!\theta\, e^{ i \beta\cos\theta}
  = \sqrt{\pi} \,\big(\fract{2}{\beta}\big)^{d-2\o 2}\GammaF{\frac{d{-}1}{2}} J_{d-2\o 2} (\beta)\,. \]
For the trace
 \[\Tr K (s,w)=\int d^{d-2}z \int_0^{2\pi\alpha} \!\!\!\! d\phi \int_0^\infty \!\! dr \,r \,K(\phi=\phi'+w,s) \]
we find that
 \begin{equation}
  \Tr K(s,w) = \frac{1}{(4\pi)^{\frac{d}{2}}} {\pi \alpha\o \sin^2{w\o 2}} A(\Sigma) P_{d-2}(s)~,
 \lb{TrK}
 \end{equation}
where $A(\Sigma )=\int d^{d-2}z$ is the area of surface $\Sigma$.
 We used the integral $\int_0^\infty dx x^{1-\nu}
J_\nu(x)={2^{1-\nu}\o \Gamma(\nu)}$ when derived (\ref{TrK}). For
the integral over contour $\Gamma$ in (\ref{Sommerfeld})
we find\footnote{
We note that convergence of integration along contour $\Gamma$ in complex $w$ plane in the Sommerfeld formula (\ref{Sommerfeld}) is not affected by modification of $F(p^2)$ compared to standard case $F(p^2)=p^2$. This can be seen by performing integration over $w$ of (\ref{Kw}) prior to integration over radial momentum $p$. The result of $w$-integration is then universal for all possible $F(p^2)$.
}
that \cite{coneQFT}, \cite{Fursaev:1994in}
 \begin{equation} \lb{C}
  C_2(\alpha)\equiv{ i \o 8\pi\alpha}\int_\Gamma \cot{w\o 2\alpha}\, {dw\o \sin^2{w\o 2}}={1\o 6\alpha^2}(1-\alpha^2)~.
 \end{equation}
Collecting everything together we obtain
 \begin{equation}
  \Tr K_\alpha(s) =\frac1{(4\pi)^{d/2}} \big( \alpha V P_{d}(s)
   \, + \,2\pi \alpha C_2(\alpha)A(\Sigma)P_{d-2}(s)\big)
 \lb{P}
 \end{equation}
for the trace of the heat kernel on space with a conical singularity.

For quadratic differential operator ${\cal D}=\Box$ one has $F(p^2)=p^2$ and $P_n(s)= s^{-n/2}$. Then (\ref{P}) reproduces the well-known result \cite{coneQFT}, \cite{Fursaev:1994in} for the Laplace operator
 \begin{equation}
  \Tr K_\alpha(s)={1\over (4\pi s)^{d/2}}\big(\alpha V + s \,2\pi\alpha C_2(\alpha)A(\Sigma)\,\big)\;,
 \lb{TrKcon}
 \end{equation}
where $V=\int d\tau d^{d-1}x$ is the volume of spacetime and $A(\Sigma)=\int d^{d-2}x$ is the area of surface $\Sigma$. Substituting (\ref{P}) into equation (\ref{W}) we obtain that the effective action contains two terms. The one proportional to the volume $V$ reproduces the vacuum energy already calculated in the previous section. The second term proportional to the area $A(\Sigma)$ is responsible for the entropy. Applying the formula (\ref{SS}) we obtain entanglement entropy
 \begin{equation}
  S=\frac{A(\Sigma)}{12\cdot (4\pi)^{(d{-}2)/2}}
  \int_{\epsilon^2}^\infty {ds\o s}P_{d-2}(s)\,.
 \lb{Sent}
 \end{equation}
The behavior of $P_{d-2}(s)$ at small $s$ depends on the behavior of function $F(p^2)$ at large momentum $p$. However, no matter how fast is function $F(p^2)$ at large $p$ the function $P_{d-2}(s)$ is always divergent in the limit $s\rightarrow 0$. Thus we conclude that no matter how good is the UV behavior of the field propagator $F^{-1}(p^2)$ the entropy (\ref{SS}) is always divergent.

The divergence however can be milder if $F(p^2)$ is growing faster than $p^2$. We consider two particular examples.
\\

\noindent{\it Example 1. $F(p^2)$ is polynomially growing function at large $p$,}
 \begin{eqnarray} \label{ex1}
  &&F(p^2)\simeq {p^{2k}\o m^{2(k-1)}}\,, \qquad
   p^2\gg m^2 \,, \\
  &&P_{d-2}(s)\simeq \frac{1}{k} \,\frac{\GammaF{\frac{d-2}{2k}}}{\GammaF{\frac{d-2}{2}}} \, \Big(\frac{m^{k-1}}{s}\Big)^{\frac{d-2}{k}} \,,\qquad
  s \rightarrow 0\,.\nonumber
 \end{eqnarray}
The UV divergence of the entropy in this case is
 \begin{equation}  \label{Sdiv}
  S_{div} = \frac{A(\Sigma)}{12 \cdot (4\pi)^{(d{-}2)/2}}\,\frac{\GammaF{\frac{d-2}{2k}}}{\GammaF{\frac{d}{2}}} \, {m^{(d-2)(k-1)\o k}\o \epsilon^{d-2\o k}}\;.
 \end{equation}
This expression has a pole at $d=2$ which indicates that in two dimensions the divergence in the entropy is logarithmic.

The interplay of the standard $p^2$ term and the UV term in the inverse propagator can be seen in $d=4$ for the function
 \begin{eqnarray} \lb{p4}
  &&F(p^2)=p^2+{p^4\o m^2}~,\\
  &&P_2(s)\simeq \frac{\sqrt{\pi}}{2} \frac{m}{\sqrt{s}}-\frac{m^2}{2}
    +O(\sqrt{s})~.\nonumber
 \end{eqnarray}
The entropy then contains a logarithmic term
 \begin{equation} \label{sp4}
  S={A(\Sigma)\o 48\pi}(\sqrt{\pi} \,\frac{m}{\epsilon} + m^2\ln(m\epsilon)+O(m^2))
 \end{equation}
as well as UV finite terms proportional to $m^2$.
\\

\noindent{\it Example 2. $F(p^2)$ is exponentially growing function at large $p$},
 \begin{eqnarray} \label{exp}
  &&F(p^2)\simeq m^2 e^{p^2/\Lambda^2}, \\
  &&P_2(s)=-{\Lambda^2\o 2}(\gamma+\ln (sm^2)+O(s)), \nonumber
 \end{eqnarray}
where we take $d=4$. The entropy in this case has only logarithmic
divergences
 \begin{equation} \lb{sexp}
  S={A(\Sigma)\o 48\pi}\, \Lambda^2\, (\ln^2(\epsilon m)+\gamma\ln(\epsilon m)).
 \end{equation}

We conclude that entanglement entropy is always UV divergent no matter how fast the function $F(p^2)$ grows. The degree of divergence however changes considerably if one considers $F(p^2)$ growing exponentially. This behavior resembles that of cosmological constant studied in Section 2.

\bigskip
\section{Newton's constant}\setcounter{equation}0
Our calculation so far has been made in flat spacetime. Now we want to make a step forward and generalize this to a weakly curved spacetimes. Although it is rather unusual to use a momentum representation in curved space we show in Appendix A that it can be  efficiently exploited on  a weakly curved gravitational background so that the first few   covariant terms in the heat kernel can be reproduced by using this method.

In Appendix A the trace of the heat kernel of operator $F(\Box_g)$  is shown to take the form
 \begin{equation} \label{4.17}
  \Tr e^{-sF(\Box_g)}= \frac1{(4\pi)^{d/2}} \left( P_{d}(s) \int d^dX\sqrt{g} +
  P_{d-2}(s) \int d^dX\sqrt{g}\: \frac16 R+ ..\right) \,,
 \end{equation}
where we omit all terms containing  more than two derivatives acting on metric. This expression generalizes the well-known small $s$ expansion of the heat kernel of the Laplace operator $\Box_g$,
 \begin{equation} \label{KKK}
  \Tr e^{-s\Box_g} = {1\o (4\pi s)^{d\o 2}}\left(\int d^dX\sqrt{g}
  + s \int d^dX\sqrt{g}\, {1\o 6}\, R+.. \right) \,.
 \end{equation}
However, we should note  that (\ref{4.17}) is a decomposition in number of derivatives of metric rather than   a small $s$ expansion. In the case of quadratic operator $\Box_g$ the two decompositions, in number of derivatives and in powers of $s$, in fact coincide.

Thus the effective action $W=-{1\o 2}\int^\infty_{\epsilon^2}{ds\o s}\Tr e^{-sF(\Box_g)}$ for a theory described by field operator $F(\Box_g)$ to lowest orders in derivative  takes the form
 \begin{equation} \lb{4.18}
  W[g_{\mu\nu}]=-\lambda(\epsilon)\int d ^dX\sqrt{g}
  -{1\o 16\pi  G_N(\epsilon)} \int d^dX \sqrt{g}\, R[g] \,,
 \end{equation}
where the induced   cosmological constant  $\lambda(\epsilon)$ is given by expression (\ref{5}) and $G_{N}(\epsilon)$ is the induced Newton's constant,
 \begin{equation} \lb{4.19}
  \frac{1}{16\pi G_{N}(\epsilon)} = \frac1{(4\pi)^{d/2}} \frac1{12}
  \int_{\epsilon^2}^\infty \!\!\frac{ds}{s} \,P_{d-2}(s) \,.
 \end{equation}

The Newton's coupling (\ref{4.19}) in the effective gravitational
action (\ref{4.18}) should be compared to the expression
(\ref{Sent}) for entanglement entropy which we have derived
earlier in this paper. The comparison shows that entanglement
entropy (\ref{Sent}) can be expressed in terms of the induced
Newton's constant (\ref{4.19}) as follows
 \begin{equation} \lb{4.20}
  S={1\o 4G_N(\epsilon)}A(\Sigma) \,.
 \end{equation}
This clearly resembles the Bekenstein-Hawking entropy of a black hole even though the
entanglement entropy (\ref{Sent}) can be defined for any surface
which should  not be necessarily a black hole horizon. The fact that there
is a precise balance between the UV divergence in entanglement
entropy and the UV divergence of Newton's constant in the
effective action was first  proposed in \cite{Susskind:1994sm},
later  this was checked in different approaches \cite{ES}, \cite{Demers:1995dq}, \cite{Solodukhin:1994yz}  and
even has been proven \cite{Fursaev:1994ea} to be a general
property of quantum fields. As the present consideration shows
this balance preserves for a more general theory which is modified
in UV by local or non-local terms in the action. This in fact
was earlier observed in \cite{Dvali:2008jb} for the entropy of
black holes in the context of the  Kaluza-Klein theories. In these
theories the propagator at short distances is essentially
higher-dimensional that manifests directly in the entropy of
microscopic black holes which shows a higher-dimensional behavior.

\bigskip
\section{Complete covariant decomposition}\setcounter{equation}0
In the previous section we used a non-covariant method and managed to extract several first covariant terms in the decomposition of the heat kernel of operator $F(\Box_g)$. Here we generalize this and in fact obtain the complete structure of the heat kernel.

First, we note that the heat kernel $e^{-\apt\Box_g}$ of operator $\Box_g$ is available in the form of a well-known expansion
 \begin{equation} \lb{c1}
  \Tr e^{-\apt\Box_g}={1\o (4\pi)^{d/2}}\sum_{n=0}^\infty \apt^{n-{d/2}}C_n \,,
 \end{equation}
where $C_n$ are the Schwinger-DeWitt coefficients \cite{DeWitt}. The first two coefficients are available in (\ref{KKK}). In general, the $n$-th coefficient is an integral of a covariant expression which contains $n$ derivatives of metric. Equation (\ref{KKK}) is thus a covariant decomposition of the heat kernel in powers of the Riemann tensor and its covariant derivatives. As we show in this section, the heat kernel of operator $F(\Box_g)$ can be decomposed in terms of exactly the same covariant combinations $C_n$. The coefficients however in this case are not simply the powers $s^{n-d/2}$ but functions of the proper time $s$ the exact form of which depends on the function $F(\Box_g)$.

We can formally rewrite the trace of $e^{-sF(\Box_g)}$ by
inserting a delta-function,
 \begin{equation} \lb{c2}
  e^{-sF(\Box_g)}= \int_0^\infty dq \,e^{-sF(q)}\,\delta(\Box_g{-}q)
 \end{equation}
and using a well-known Fourier form for the delta-function, we find a very useful representation
\footnote{This representation is valid if there exists one-to-one correspondence between eigenvalues of operators $F(\Box_g)$ and $\Box_g$. For manifolds without boundaries this condition is satisfied.}
 \begin{equation} \lb{Formula}
  e^{-sF(\Box_g)}
  = \int_0^\infty dq\, e^{-sF(q)}{1\o 2\pi i} \int_{-i\infty}^{+i\infty}d\apt \,e^{q\apt} \,e^{-\Box_g\apt} \,,
 \end{equation}
\begin{wrapfigure}{l}{0.5\textwidth}
  \vspace{-20pt}
  \begin{center}
    \includegraphics[width=0.33\textwidth]{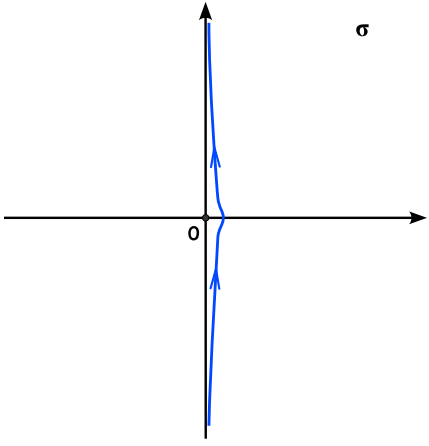}
  \end{center}
  \vspace{-20pt} \label{fig1}
  \caption{\textit{Contour of $\apt$-integration starts at negative imaginary infinity, ends at positive imaginary infinity and passes zero from the positive real side.\newline }}
\end{wrapfigure}
which reduces the heat kernel of operator $F(\Box_g)$ to the heat kernel of the well studied and much simpler second order differential operator $\Box_g$. Contour of integration over $\apt$ is shown in Figure~1. Its form is dictated by condition of convergence.
Since the expansion (\ref{c1}) can be analytically
continued to complex values of $\apt$ (provided it has a positive real part) we arrive at the following expression for the heat kernel
\begin{equation} \lb{c3}
  \Tr e^{-sF(\Box_g)}={1\o (4\pi)^{d/2}}\sum_{n=0}^\infty {\cal T}_n(s)C_n \,,
 \end{equation}
where
 \begin{eqnarray} \lb{ss}
  &&\!\!\!\!\!\!\!\!{\cal T}_{n}(s)=\int_0^\infty dq\, e^{-sF(q)} {\cal A}_n(q) \,,
 \\ \lb{c4}
  &&\!\!\!\!\!\!\!\!{\cal A}_n(q)={1\o 2\pi i}\int_{-i\infty}^{+i\infty}d\apt \,\apt^{n-d/2} e^{q\apt}.
 \end{eqnarray}

Evaluating the $\apt$-integral in (\ref{c4}) we should consider two cases:
\\
$\phantom{.}\qquad$ i) $n\geq d/2$, the corresponding terms in expansion (\ref{c1}) are regular terms;
\\
$\phantom{.}\qquad$ ii) $n<d/2$, the corresponding terms in the expansion (\ref{c1}) are singular.
\\

First, we consider the case of regular terms, $n\geq d/2$, and additionally assume that $d/2$ is an integer. Then the integral in (\ref{c4}) can be presented in the form
 \begin{equation}
  {\cal A}_n(q)=
  \partial_q^{n-d/2} \left( {1\o 2\pi i} \int_{-i\infty}^{+i\infty}\!\!\!\!d\apt\, e^{q\apt}\right)=\partial_q^{n-d/2}\delta(q)
 \end{equation}
and hence the corresponding integral over $q$ in (\ref{ss}) reduces to\footnote{If $d/2$ is not integer the $q$-derivative in (\ref{c5}) should be understood as a fractional derivative.}
 \begin{equation} \lb{c5}
  {\cal T}_n=(-\partial_q)^{n-{d/2}}e^{-sF(q)}\Big|_{q=0} \,.
 \end{equation}

For $n<d/2$ the contour in the integral (\ref{c4}) can be deformed
to encircle counter clock-wise the negative real axis. The
$\tau$-integration then gives a $\Gamma$-function (see Eq.(8.315)
in \cite{GradRyzh}),
 \begin{equation} \lb{c6}
  {\cal A}_n(q) = {q^{d/2-n-1}\o \GammaF{d/2{-}n}} \,.
 \end{equation}
 The function (\ref{ss}) in this case
is thus represented by the integral
 \begin{equation} \lb{c7}
  {\cal T}_{n}(s) = \frac{1}{\GammaF{d/2{-}n}}\int_0^\infty dq \,q^{d/2-n-1 }e^{-sF(q)}
                  = P_{d-2n}(s) \,,
 \end{equation}
where functions $P_k(s)$ are those defined in (\ref{4}).

Collecting everything together we find that the heat kernel $\Tr e^{-sF(\Box_g)}$ has the following covariant decomposition
 \begin{equation}  \lb{c8}
  \Tr e^{-sF(\Box_g)}={1\o (4\pi)^{d/2}}\sum_{n=0}^\infty {\cal T}_n(s)\,C_n  \,,
 \end{equation}
where $C_n$ are the Schwinger-DeWitt coefficients for operator $\Box_g$ \cite{DeWitt} and  the functions ${\cal
T}_n(s)$ are defined as follows
 \begin{equation} \lb{c9}
  {\cal T}_n(s) = \left\{
              \begin{array}{ll}
                 P_{d-2n}(s)
                  &  n<d/2 \\
                (-\partial_q)^{n-d/2} e^{-s F(q)}\Big|_{q=0}
                  &  n\geq d/2
              \end{array}
    \right.
 \end{equation}
The singular terms  $n=0$ and $n=1$  in equations (\ref{c8}), (\ref{c9}) agree with the expression (\ref{4.17}) obtained in the previous section.

It should be noted that as can be seen from (\ref{c9}) in general the total number of singular terms in the heat kernel and in the corresponding effective action is preserved under a Lorentz invariant deformation of operator $\Box_g$. In fact, an UV divergent term in a theory with operator $\Box_g$ remains to be divergent (although with a different degree of divergence) in a theory with operator $F(\Box_g)$. The logarithmically divergent term  for $n={d/2}$ in a theory with $\Box_g$ still diverges logarithmically in a theory with $F(\Box_g)$.

Clearly, our result (\ref{c8}), (\ref{c9}) generalizes to the heat kernel of any operator which is a function of a differential  operator for which the trace of the  heat kernel is known in a form similar to (\ref{c1}).

When function $F(p^2)$ is growing not faster than $p^2$ for large values of $p$ as in (\ref{i6}), (\ref{i7}) we may deform the contours of integration in (\ref{Formula}) in a way that it becomes a Laplace type transform
 \be
  e^{-sF(\Box_g)}= {1\o 2\pi i}  \int_{-i\infty}^{+i\infty}\!\! dq\, e^{-sF(q)}\int_0^\infty d\apt\, e^{q\apt}e^{-\Box_g\apt} \,.\lb{Laplace}
 \ee
The trace  of this heat kernel still takes the form (\ref{c8}), (\ref{c9}). Sometimes it may be more convenient to use the Laplace type representation
since the  heat kernel then is expressed in terms of heat kernel $e^{-\sigma\Box_g}$ where $\sigma$ is real variable.

\bigskip

\section{Non-Lorentz invariant UV modifications}\setcounter{equation}0
As has been understood recently (see \cite{Horava}, \cite{Son}, \cite{Balasubramanian:2008dm} and citations to these papers for the latest development) the UV modifications should not be necessarily  Lorentz invariant. In fact, by changing the theory in UV in a non-Lorentz invariant fashion one can get a renormalizable theory which is free of the ghosts. The gravitational degrees of freedom still can be represented as a metric
 \begin{equation} \lb{5.1}
  ds^2=N^2(\tau,x) d\tau^2+2N^i(\tau,x) d\tau dx^i+g_{ij}(\tau,x) dx^i dx^j \,.
 \end{equation}
In this paper we use the gauge in which  $N^i(\tau,x)=0$. The time coordinate $\tau$ and the spatial coordinates $\{ x^i,\, i=1,..,d{-}1\}$ then appear in a different way in the field operator
 \begin{equation} \lb{5.2}
  {\cal D}=\Delta_\tau+F(\Delta_x) \,,
 \end{equation}
where
 \[ \Delta_\tau=-{1\o N\sqrt{g}}\partial_\tau ({\sqrt{g}\o N}\partial_\tau ) \]
and
 \[ \Delta_x=-{1\o N\sqrt{g}}\partial_{i}(N\sqrt{g}g^{ij}\partial_j) \,. \]
We remind the reader that throughout this paper we work in the Euclidean signature.  Operator (\ref{5.2}) is a curved spacetime version of the Lifshitz operator \cite{Lifshitz} used to describe a phase transition in condensed matter.

The quantum theory described by operator (\ref{5.2})  is still characterized by an effective action which is a functional of the background gravitational field,
 \begin{equation}
  W[N,g_{ij}]=-{1\o 2}\int_{\epsilon^2}^\infty {ds\o s}\; \Tr  e^{-s{\cal D}} \,.
 \lb{5.3}
 \end{equation}
As in the Lorentz invariant case the effective action can be decomposed according to the
number of derivatives of the gravitational field. To  lowest
order, restricting to the number of derivatives not higher than
two, we get\footnote{The other possible terms quadratic in
derivatives are integrals of $(\partial_\tau N)^2$ and $K \partial_\tau{N}$. These
terms however should be excluded as they are not invariant under
the transformation $\tau=\tau(\tau'),~N(\tau)=N(\tau'){\partial
\tau'\o \partial \tau}$ which is a symmetry of operator
(\ref{5.2}). }
 \begin{equation} \lb{5.4}
  W=
  \int d\mu \left(-\lambda(\epsilon)+b_1(\epsilon) R^{(d-1)} +b_2(\epsilon) k_{ij}^2+b_3(\epsilon) k^2+b_4(\epsilon) {(\nabla N)^2\o N^2}\right)~,
 \end{equation}
where $R^{(d-1)}$ is the Ricci scalar for the spatial $(d-1)$-metric $g_{ij}$ and we introduced
the measure
 \begin{equation} \lb{5.5}
  d\mu=d\tau d^{d-1}x N\sqrt{g}
 \end{equation}
and the extrinsic curvature
 \begin{equation} \lb{5.6}
  k_{ij}=\frac{1}{2N}\partial_\tau {g}_{ij} \,, \qquad
  k=g^{ij}k_{ij} \,.
 \end{equation}
The covariant derivative $\nabla$ is determined with respect to metric $g_{ij}$. The couplings $b_j,~j=1,..,4$ are now the gravitational couplings which replace the single coupling, Newton's constant, in the Lorentz invariant case. $\lambda(\epsilon)$ still has the meaning of the vacuum energy (or cosmological constant).
In this section we evaluate explicitly $\lambda(\epsilon)$ and couplings $b_j(\epsilon),\, j=1,..,4$ that depend on the UV cutoff $\epsilon$. This calculation is rather involved. That is why it is useful to start with a simple consideration which uses some scaling arguments.

\bigskip
\subsection{Scaling arguments}
Consider a quantum field described by a field operator of the type
(\ref{5.2}) where the function $F$ is a power of
spatial Laplace operator $\Delta_x$,
 \begin{equation} \lb{5.7}
  F(\Delta_x)={(\Delta_x)^{k}\o m^{2(k-1)}} \,.
 \end{equation}
The heat kernel $e^{-s{\cal D}}$ then is invariant under the scaling transformations
\cite{Solodukhin:2009sk}
 \begin{equation} \lb{5.8}
  \tau\rightarrow \beta \tau \,,
  \qquad x^i\rightarrow \beta^{1\o k}x^i \,,
  \qquad s\rightarrow \beta^2 s \,.
 \end{equation} Under these transformations the metric
invariants transform as
 \begin{equation} \lb{5.9}
  R^{(d-1)}\rightarrow \beta^{-{2\o k}}R^{(d-1)} \,, \qquad
  k_{ij}\rightarrow \beta^{-1}k_{ij} \,.
 \end{equation}
Since the effective action (\ref{5.4}) is invariant under (\ref{5.8}) (provided that $\epsilon\rightarrow \beta\epsilon$) we immediately obtain   the dependence on $\epsilon$ for the pre-factors that appear in (\ref{5.4})
 \begin{eqnarray} \lb{5.10}
  &&\lambda(\epsilon)\sim {1\o \epsilon^{d+k-1\o k}} \,,
   \qquad b_1(\epsilon)\sim b_4(\epsilon)\sim {1\o \epsilon^{d+k-3\o k}} \,,
  \nonumber \\
  &&b_2(\epsilon)\sim b_3(\epsilon)\sim  {1\o \epsilon^{d-1-k\o k}} \,.
\end{eqnarray}
The dependence on the mass parameter $m$ of the coefficients (\ref{5.10}) can be easily restored by dimensional analysis.

We remind that in the Lorentz invariant case there was a precise balance due to the Gauss-Codazzi relations, $b_1=b_3=-b_2$. This balance  does not survive if the Lorentz symmetry is broken as we see from (\ref{5.10}). The interesting fact is that the factors $a(\epsilon)$, $b_1(\epsilon)$  and $b_4(\epsilon)$  are always divergent no matter what is the value of the parameter $k$ in (\ref{5.7}). On the other hand, the extrinsic curvature  terms in
the effective action are \emph{UV finite} if
 \begin{equation}  \lb{k}
  k>d-1 \,.
 \end{equation}
(For $k=d-1$ these terms are logarithmically divergent.) In four spacetime dimensions the critical value for $k$ is 3. This seems to be consistent with the proposal of Horava \cite{Horava} for the UV modification of gravity action.

One can analyze the UV divergence of more general terms that contain powers of extrinsic curvature and powers of the curvature of $(d{-}1)$-metric $g_{ij}$,
 \begin{equation} \lb{KR}
  {1\o \epsilon^\varkappa}\int d\mu \,{\cal K}^{2p}{\cal R}^n \,,
 \end{equation}
where ${\cal K}$ stands for components of extrinsic curvature and $\cal R$ stands for the components of the Riemann curvature of metric $g_{ij}$. The degree of divergence of the term (\ref{KR}) is
 \begin{equation} \lb{lambda}
  \varkappa={d-1+k-2n-2pk\o k}~.
 \end{equation}
If  $p\ge 1$ we find that all terms of the form (\ref{KR}) that contain extrinsic curvature are UV finite if
$k$ satisfies the condition (\ref{k}). On the other hand, if (\ref{KR}) contains only the Riemann curvature, i.e.  $p=0$, the term (\ref{KR}) is UV divergent when
  \[ n\leq {d-1+k\o 2} \,.\]
Comparing to the Lorentz invariant case, $k=1$,  we find that the theory with a broken Lorentz invariance contains more UV divergent terms. For instance, the term with $n=3$ in dimension $d=4$ is UV finite in the Lorentz invariant case ($k=1$) while it is logarithmically divergent if the theory is perturbed by operator (\ref{5.2}) with $k=3$.

Thus, we observe an interesting phenomenon: some UV finite terms in the Lorentz invariant theory become UV divergent if the theory is perturbed by a non-Lorentz invariant operator.

Provided condition (\ref{k}) is satisfied we obtain that the gravitational theory in UV contains no terms with time derivatives. Up to the second order in derivative the theory in UV regime is described by the action
 \begin{equation} \lb{UV}
  W_{UV}=\int d\tau d^{d-1}x\sqrt{g} N \left(-\lambda + b_1 R^{(d-1)}+b_4 {(\nabla N)^2\o N^2}\right).
 \end{equation}

By the same scaling arguments we find the structure of entanglement entropy in a theory with the field operator (\ref{5.2}). The scaling law for the area $A(\Sigma)=\int d^{d-2} x$ is the following
  \[ A\rightarrow \beta^{d-2\o k}A \,. \]
Hence we find that \cite{Solodukhin:2009sk}
 \begin{equation}
  S\sim {A(\Sigma)\o \epsilon^{d-2\o k}} \,. \lb{5.11}
 \end{equation}
We immediately note that the predicted UV divergence of the entropy is the same as in the Lorentz invariant theory (see Eq. \ref{Sdiv}). Earlier in \cite{Solodukhin:2009sk} it was shown that the UV divergences of entanglement entropy in a non-relativistic theory described by the Schroedinger type wave operator are the same as in the relativistic theory. As the scaling arguments show, this observation may be valid also for a more general type of theories with broken Lorentz invariance. In the following sections we demonstrate this in an explicit calculation of the entropy.

As we have seen in the previous sections,  in the Lorentz invariant case the entanglement entropy is associated with a particular term in the effective action so that the UV divergence of the entropy is identical to the UV divergence of the Newton's constant. As a simple comparison of (\ref{5.11}) and (\ref{5.10}) shows the UV divergence (\ref{5.11}) is not anymore associated with any term in the effective action (\ref{5.4}). Thus the balance between the gravitational action and the entropy does not take place if the Lorentz invariance is broken.

\bigskip

\subsection{Entanglement entropy: non-Lorentz invariant theories}
In this section for simplicity we use a flat space approximation, $N(\tau,x)=1$, $g_{ij}(\tau,x)=\delta_{ij}$. So that in (\ref{5.2}) two operators $\Delta_\tau=-\partial^2_\tau$ and $\Delta_x=-\partial_i^2$ commute.
As in the Lorentz invariant case computing entanglement entropy associated with a surface $\Sigma$ we choose  $(d-1)$ spatial coordinates $\{x^i,~i=1,..,d-1\}=\{x,z^a, ~a=1,..,d-2\}$, where $x$ is the coordinate orthogonal to the surface $\Sigma$ and $z^a$ are the coordinates on the surface $\Sigma$. Then we switch to the polar coordinates, $\tau=r\sin(\phi)$, $x=r\cos(\phi)$.  In the Lorentz invariant case  the conical space which is needed for calculation of entanglement entropy is obtained by making the angular coordinate $\phi$ periodic with the period $2\pi \alpha$ by applying the Sommerfeld formula (\ref{Sommerfeld}) to the heat kernel.

If the Lorentz invariance is broken, as it is for the operator ${\cal D}=\Delta_\tau+F(\Delta_x)$, there are  certain difficulties in applying the method of the conical singularity when one computes the entanglement entropy.
The difficulties come from the fact that the wave operator $\cal D$, if written in terms of the polar coordinates $r$ and $\phi$, becomes an explicit function of the angular coordinate $\phi$. As a result of this the operator $\cal D$ is not invariant under shifts of $\phi$ to an arbitrary constant $\phi+w$.  Only shifts with $w=2\pi n$ where $n$ is integer are allowed. On one hand, this is a good news since a conical space with angle deficit $2\pi (1-n)$ is exactly what we need to compute $\Tr \rho^n$ for the reduced density matrix (\ref{W2}). The bad news however is that in this case one cannot apply the Sommerfeld formula since it explicitly uses the symmetry of the differential operator under shifts of angle $\phi$.

The formal solution to the heat equation for operator $\cal D$ with required $2\pi n$ periodicity exists in the form
 \begin{equation} \lb{formal}
  e^{-s{\cal D}}\, {1\o r}\delta(r-r') \sum_{m\in Z}\delta(\phi-\phi'+2\pi n m)\,.
 \end{equation}
However, we did not manage to find an explicit coordinate representation for the heat kernel (\ref{formal}) which may be useful to our purposes. Even more, we suspect that the action of the operator $\cal D$ on the delta-function in (\ref{formal}) may be ill-defined at the location of the tip of the cone, $r=0$.

Since the direct computation is at the moment not available we have to use indirect reasoning and scaling arguments.
We use a representation similar to (\ref{Formula}),
 \begin{equation} \lb{E1}
  \Tr e^{-s\Delta_\tau-sF(\Delta_x)} = \int_0^\infty dq \;
  e^{-sF(q)}{1\o 2\pi i}\int_{-i\infty}^{i\infty}d\sigma \;
  e^{q\sigma}\; \Tr e^{-(s\Delta_\tau+\sigma \Delta_x)}~.
 \end{equation}
The trace of the heat kernel on a conical space with $2\pi n$ periodicity then reduces to the trace  of exponential of  operator $(s\Delta_\tau+\sigma\Delta_x)$ (we note that $\sigma$ is a complex number). The conjectured form for this trace is
 \begin{equation} \lb{trace}
  \Tr_n e^{-(s\Delta_\tau+\sigma\Delta_x)}={1\o (4\pi s)^{1\o 2}}{1\o (4\pi\sigma)^{d-1\o 2}}\left(a(n)V+b(n)\sqrt{s\sigma}A(\Sigma)\right)~,
 \end{equation}
where
 \begin{equation} \lb{ab}
  a(n)=n\,,\quad b(n)=2\pi nC_2(n)~,
 \end{equation}
and $C_2(n)$ is defined in (\ref{C}).

\bigskip

Below we give several arguments which support our conjecture:

\bigskip

i) If both $s$ and $\sigma$ are real we can redefine the time coordinate $\tau'=\sqrt{s\over \sigma}\tau$. Then we have that
$s\Delta_\tau+\sigma\Delta_x=\sigma \tilde{\Box}$, where $\tilde{\Box}$ is Laplace operator written in terms of coordinates $(\tau',x)$.
Denoting by $(r',\phi')$ the corresponding polar coordinates we find a relation between two systems of polar coordinates
\be
\tan\phi'=\sqrt{s\over \sigma}\tan\phi~,~~r'=r\sqrt{{s\over \sigma}\sin^2\phi+\cos^2\phi}~.
\lb{rphi}
\ee
The eigenfunctions of operator $\tilde \Box$ regular at the origin and  periodic with period  $2\pi n$ in angular coordinate $\phi'$ are easily constructed (see for example \cite{Deser:1988qn}),
$J_{|m|\over n}(\lambda r') e^{im\phi'\over n}$. Clearly, these eigenfunctions if expressed in terms of original polar coordinates $(r,\phi)$ remain regular in the origin and periodic with period $2\pi n$ in angular coordinate $\phi$. The latter follows from the fact that relation (\ref{rphi}) between two angular coordinates is $2\pi$ periodic and hence a $2\pi n$ periodic function of $\phi'$ is automatically a $2\pi n$ periodic function of $\phi$. We conclude that the trace of $e^{-(s\Delta_\tau+\sigma\Delta_x)}$ on a conical space can be identified with the heat kernel trace of operator $\tilde \Box$ on the same space. The latter is well-known and given by (\ref{TrKcon}). Making the coordinate transformation $\tau'=\sqrt{s\over \sigma}\tau$ in (\ref{TrKcon}) we arrive at (\ref{trace}), (\ref{ab}).

\bigskip

ii) If $\sigma$ is complex the above procedure does not seem to work since $r'$ becomes a complex variable so that $J_{|m|\over n}(\lambda r')$ in the eigenfunction becomes divergent for large values of $r$ for certain values of $\phi$ (for instance $\phi=\pi$). However, we note that in the general case of complex $s$ and $\sigma$ the operator $(s\Delta_\tau+\sigma\Delta_x)$ is invariant under independent rescaling of $\tau$ and $x^i$:
 \begin{eqnarray} \lb{trans}
  &&\tau\rightarrow \nu\tau~,~~s\rightarrow \nu^2 s~,\nonumber \\
  &&x^i\rightarrow \mu x^i~,~~\sigma\rightarrow \mu^2 \sigma~.
 \end{eqnarray}
Computing the trace of $e^{-(s\Delta_\tau+\sigma\Delta_x)}$ on space with $2\pi n$ periodicity we expect that there are two contributions to the trace: one from the volume and the other from the area of the surface $\Sigma$. Then the general structure of the trace is fixed by the symmetry (\ref{trans}) and is given by (\ref{trace}).
If $s=\sigma$ we note that $\Delta_\tau+\Delta_x$ is the Laplace operator. The trace of the heat kernel in this case is given by (\ref{TrKcon}). This fixes $a(n)$ and $b(n)$ as in (\ref{ab}).

\bigskip

iii) If function $F(q)$ grows slower than $q$ we can use a Laplace transform  (\ref{Laplace}) to express $e^{-sF(\Delta_x)}$ in terms of $e^{-\sigma\Delta_x}$, where now $\sigma$ is real variable. The exact form of the Laplace transform can be obtained by deforming the contours of integration over $q$ and $\sigma$ in (\ref{E1}). In this case the arguments of i) apply and we compute the trace of $e^{-(s\Delta_\tau+\sigma\Delta_x)}$ as in i). Deforming the contours of integration over $q$ and $\sigma$ back as in (\ref{E1}) we again obtain (\ref{trace}), (\ref{ab}).

\bigskip

We consider this set of indirect arguments as  sufficient to support our conjectured form (\ref{trace}), (\ref{ab}).

Taking (\ref{trace}), (\ref{ab}) and performing  integration over $\sigma$ and $q$ in (\ref{E1}) we find for
the trace of the heat kernel $K(s)=e^{-s{\cal D}}$ (\ref{E1}) on a conical space with $2\pi n$ periodicity,
 \begin{equation} \lb{E4}
  \Tr K_n(s)=n\Tr K_{n=1}(s)+\frac{1}{(4\pi)^{d/2}} 2\pi n\,C_2(n)\,A(\Sigma)\,P_{d-2}(s) \,,
 \end{equation}
where $n\Tr K_{n=1}(s)$ is the bulk contribution. By the arguments presented in Section 3.1 there is unique analytic extension of this formula to non-integer $n$. A simple comparison with the surface term in the heat kernel  of the
Lorentz invariant operator which was obtained in Section 3 shows that the surface terms of  two kernels are identical. We thus conclude that entanglement entropy is given by the same formula
 \begin{equation} \lb{E6}
  S=\frac{A(\Sigma)}{12\cdot (4\pi)^{(d{-}2)/2}}
  \int_{\epsilon^2}^\infty {ds\o s}P_{d-2}(s)
 \end{equation}
as in the Lorentz invariant case  (\ref{Sent}). Earlier a similar property of entanglement entropy was observed for a non-relativistic theory described by the Schroedinger operator \cite{Solodukhin:2009sk}. Here we see that this property is valid for a generic theory with the broken Lorentz symmetry. The fact that entanglement entropy is the same in a relativistic theory and in a theory with the broken Lorentz symmetry, provided the function $F$ is the same in both cases, is surprising. This means that the entropy, although  sensitive to the UV modifications of the theory, does not distinguish whether these modifications are Lorentz invariant or not\footnote
{We are aware of the results \cite{GW} indicating that entanglement entropy of free fermions may scale faster than area law. Apparently this is due to the presence of the Fermi surface in these models. It would be nice to understand whether these systems can be studied with our methods.}.

\bigskip
\subsection{Gravitational couplings in curved spacetime}
Our method based on equation (\ref{Formula})  that  we have used in flat spacetime is applicable in the
case of a ``curved space-time'', when the lapse function $N$ and
the $(d-1)$-metric $g_{ij}$ are arbitrary functions of $\tau$ and
$x^i$, $i{=}1,..\,,d{-}1$. The main technical difficulty, however, is that the two
operators, $\Delta_\tau$ and $\Delta_x$, do not commute in this
case and, in particular, $e^{-s(\Delta_\tau+F(\Delta_x))}\neq
e^{-s\Delta_\tau}\cdot e^{-sF(\Delta_x)}$. So that a more general
formula, known as the Zassenhaus formula should be used (see Appendix C). With this
technique applied step by step the calculation of the heat kernel trace $\Tr K(s)=\Tr e^{-s(\Delta_\tau+F(\Delta_x))}$ is
straightforward but tedious. An essential simplification is that
we are interested only in the terms which are quadratic in
derivatives of the ``metric'': the lapse function $N$ and metric
$g_{ij}$. Details of the calculation are collected in Appendix
C. Here we just present the final result
  \begin{equation}\label{7.1}
   \Tr K(s)= \int d\mu \left(A_0(s) +{B}_1(s)R^{(d{-}1)}+B_2(s) k_{ij}^2+ B_3(s) k^2
                         +  B_4(s) {(\nabla N)^2\o N^2}+..\right)\,
  \end{equation}
where $d\mu=d\tau d^{d{-}1}xN\sqrt{g}$ is a covariant measure,
$R^{(d{-}1)}$ is the curvature for $(d{-}1)$-metric $g_{ij}$,
$\nabla$ is the covariant derivative
defined with respect to $g_{ij}$, and we skip all terms with more
than two derivatives. The exact expressions for the functions of
proper time are
  \begin{eqnarray}  \label{7.2}
   & A_0(\spt)&= \frac{1}{(4\pi)^{d/2}}\, \spt^{{-}1/2}\, P_{d-1}(\spt)\;,\\
   & B_1(\spt)&= \frac{1}{(4\pi)^{d/2}}\,  \spt^{{-}1/2}\, \frac16 \,P_{d-3}(\spt)\;,
     \nonumber\\
   & B_2(\spt)&= \frac{1}{(4\pi)^{d/2}}\,  \spt^{{-}1/2}\, \Big(
     {-}\frac{\spt }6\, P_{d-1}(\spt) - \frac{\spt^2}6\, \widetilde{P}_{d+3}(\spt)\Big)\;,
     \nonumber \\
   & B_3(\spt)&= \frac{1}{(4\pi)^{d/2}}\, \spt^{{-}1/2}\,
      \Big(\frac{\spt }6\, P_{d-1}(\spt) - \frac{\spt^2}{12} \widetilde{P}_{d+3}(\spt)\Big)\;,
     \nonumber\\
   & B_4(\spt)&= \frac{1}{(4\pi)^{d/2}}\,  \spt^{{-}1/2}\,
     \Big({-}\frac{\spt }4 \widetilde{P}_{d+1}(s)\Big)\;,
     \nonumber
  \end{eqnarray}
where we used functions $P_{n}(s)$ defined earlier (\ref{4}) and introduced $\widetilde{P}_{n}(s)$ :
 \begin{equation}
  P_{n}(s)=\frac1{\GammaF{\frac{n}2}} \int_0^\infty dq\, q^{\frac{n}2-1}\, e^{-sF(q)}\,,\qquad
  \widetilde{P}_{n}(s)=\frac1{\GammaF{\frac{n}2}} \int_0^\infty dq\, q^{\frac{n}2-1}\,F''(q)\, e^{-sF(q)}\,.
 \label{HLFormFactors}
 \end{equation}
Here we note that in particular case $F(\pBx)=\pBx$ when Lorentz invariance restores terms $\widetilde{P}_{n}(s)$ vanish and terms ${P}_{n}(s)$ give needed powers of $\spt$ that reproduces (\ref{KKK}).

The UV divergent terms in the effective action (\ref{5.4}) are expressed in terms of the functions (\ref{7.2})
  \begin{eqnarray}
  &&\lambda(\epsilon)={1\o 2}\int_{\epsilon^2}^\infty {ds\o
   s}A_0(s)~,\nonumber\\
  &&b_k(\epsilon)=-{1\o 2}\int_{\epsilon^2}^\infty {ds\o
   s}B_k(s)~,~~k=1,\, 2,\, 3,\, 4\,.
 \lb{7.3}
  \end{eqnarray}
As follows from (\ref{7.2}) the functions $A_0(s)$ and $B_1(s)$ are always
positive and respectively the cosmological constant
$\lambda(\epsilon)$ and the coupling $b_1(\epsilon)$ in the
effective action (\ref{5.4}) are negative. If the function $F(q)$
is such that $F''(q)\leq 0$ then the function $B_2(s)$ is negative
and respectively the coefficient $b_2(\epsilon)$ in front of
$k_{ij}^2$  in the effective action (\ref{5.4}) is positive. The
sign of $B_3(s)$ and respectively $b_3(\epsilon)$ depends on the
form of the function $F(q)$.

It is useful to analyze the coefficients (\ref{7.2}) for function $F(q)=q+m^{2-2k}q^k$. There are two regimes to consider. In the first regime $s\gg 1/m^2$. This could be called an infra-red regime since in this case one can neglect the higher momentum term in $F(q)$ and approximate it by the linear term $F(q)\simeq q$, the resulting theory then is Lorentz invariant and the coefficients $A_0(s)$ and $B_n(s)$ are the same  as for the relativistic Laplace operator,
 \begin{eqnarray}
  &&A_0(s)={1\o (4\pi s)^{d\o 2}}(1+O({1\o sm^2}))~,
  ~~B_4(s)=O({1\o sm^2})~,
  \nonumber\\
  &&B_1(s)=B_3(s)=-B_2(s)={1\o (4\pi s)^{d\o 2}}\;{s\o 6}(1+O({1\o sm^2}))~.
 \end{eqnarray}
This is the regime where the Lorentz invariance in the effective action (\ref{5.4}) restores.

In the other regime, $s\ll 1/m^2$, one can neglect the linear in
$q$ term and approximate $F(q)\simeq m^{2-2k}q^k$. This is
essentially the UV regime.  In this regime we find the approximate
values for the coefficients $A_0(s)$, $B_n(s)$
 \begin{eqnarray}
      &\tilde{A}_0({s})&\simeq
                    m^{(d-1){(k-1)\o k}} {s}^{-\frac{d-1}{2k}} \; \frac{\Gamma{(\frac{d-1}{2k})}}{k\Gamma{(\frac{d-1}{2})}}\;,
      \nonumber\\
      & \tilde{B}_1({s})&\simeq
                   \frac16 m^{(d{-}3)(\frac{k{-}1}{k})} {s}^{-\frac{d-3}{2k}} \; \frac{\Gamma{(\frac{d-3}{2k})}}{k\Gamma{(\frac{d-3}{2})}}\;,
      \nonumber\\
      & \tilde{B}_2({s})&\simeq
                    -\frac{1 }6  m^{(d-1)(\frac{k{-}1}{k})} {s}^{-\frac{d-1}{2k}+1}\left[ \; \frac{\Gamma{(\frac{d-1}{2k})}}{k\Gamma{(\frac{d-1}{2})}}
                    + \; \frac{(k{-}1)\Gamma{(\frac{d+3}{2k})}}{\Gamma{(\frac{d+3}{2})}}
                    \right]\;,
      \nonumber\\
      & \tilde{B}_3({s})&\simeq \frac{1 }6  m^{(d-1)(\frac{k{-}1}{k})} {s}^{-\frac{d-1}{2k}+1}
                    \left[  \; \frac{\Gamma{(\frac{d-1}{2k})}}{k\Gamma{(\frac{d-1}{2})}}
                     -
                     \frac{1}{2} \; \frac{(k{-}1)\Gamma{(\frac{d+3}{2k})}}{\Gamma{(\frac{d+3}{2})}}
                    \right]\;,
      \nonumber\\
      & \tilde{B}_4({s})&\simeq -
                    \frac{1 }4  m^{(d{-}3)(\frac{k{-}1}{k})} {s}^{{-}\frac{d-3}{2k}  } \; \frac{(k{-}1)\Gamma{(\frac{d+1}{2k})}}{\Gamma{(\frac{d+1}{2} )}}\,.
 \label{7.4}
 \end{eqnarray}
where we introduced $\tilde{A}_0(s)=(4\pi)^{d\o 2}s^{1/2}A_0(s)$ and $\tilde{B}_n(s) =(4\pi)^{d\o 2} s^{1/2}B_n(s)$.

The two regimes, the infra-red and the UV, are important when we analyze the UV divergences in the gravitational action (\ref{5.4}). This can be illustrated for the cosmological constant $\lambda(\epsilon)$. The UV regularization parameter $\epsilon$, although supposed to be small, can be large or small as compared to the mass parameter $m$ which sets the scale at which the UV modification of the field operator becomes important.  Thus, if $\epsilon\gg 1/m$ we are in the regime when the Lorentz symmetry is restored and the cosmological constant takes the usual form
 \begin{equation} \lb{7.6}
  \lambda_{IR}(\epsilon)={1\o d(4\pi)^{d\o 2}}{1\o \epsilon^d} \,.
 \end{equation}
On the other hand, in the UV regime, when $\epsilon\ll 1/m$, the cosmological constant  is
 \begin{equation} \lb{7.7}
  \lambda_{UV}(\epsilon)={1\o (4\pi)^{d\o 2}}{k\o (d+k-1)} {m^{(d-1){(k-1)\o k}}\o \epsilon^{d+k-1\o k}} {\Gamma({d-1\o 2k})\o  k \Gamma({d-1\o 2})} \,.
 \end{equation}

The most interesting behavior is shown by the couplings $b_2(\epsilon)$ and $b_3(\epsilon)$ in front of the extrinsic curvature in the effective action (\ref{5.4}). In the infra-red regime, these couplings scale as
 \[ b^{IR}_{2(3)}(\epsilon)\sim {1\o \epsilon^{d-2}} \]
and may seem divergent. However, in the UV regime, when $\epsilon\ll 1/m$, the coupling $b_2$ ($b_3$) takes a finite value if the UV term in the field operator is characterized by $k>d-1$.  This is an interesting example of
how some UV divergences can be cured by the higher momentum modifications in the propagator. The UV value of the coefficients are calculated explicitly
 \begin{eqnarray}
 &&b^{UV}_2={1\o 24(4\pi)^{d/2}}{((d^2-1)k-d^2+2d-3)\o (k-1)^2}\;{\Gamma ({k-d+1\o 2k-2})\Gamma({d-2\o 2k-2})\o \Gamma({d+3\o 2})} \,,
 \nonumber \\
 &&b^{UV}_3=-{1\o 24(4\pi)^{d/2}}{((d^2-1)k-d^2+3-d)\o (k-1)^2}\;{\Gamma ({k-d+1\o 2k-2})\Gamma({d-2\o 2k-2})\o \Gamma({d+3\o 2})} \,.
 \lb{7.8}
 \end{eqnarray}
So that the ratio of two couplings is
 \begin{equation} \lb{7.9}
  {b^{UV}_2\o b^{UV}_3}=-{(d^2-1)k-d^2+2d-3\o(d^2-1)k-d^2+3-d} \,.
 \end{equation}
It is curious that this ratio becomes $-1$ in the limit of infinite $k$ so that the balance between the two extrinsic curvature terms in the effective action (\ref{5.4}) is the same as in the General Relativity.

\bigskip

\section{Conclusions}
In conclusion let us list our main results

\begin{itemize}
	\item  We have computed the induced cosmological constant and Newton's constant in a Lorentz invariant theory with propagator $G(p^2)=1/F(p^2)$, \;$F(p^2)=p^2+f(p^2)$, where $f(p^2)$ is growing faster than $p^2$ and analyzed the dependence of these constants on the UV cutoff $\epsilon$. We have found that generically there are two regimes: in the infra-red regime, when $\epsilon$ is sufficiently large the induced constants are effectively determined by  $p^2$ term in the inverse propagator and take the standard form in terms  of $\epsilon$. In the UV regime the modification $f(p^2)$ becomes important and it changes the degree of divergence of the quantities.
	\item We have observed that if function $f(p^2)$ is exponentially growing, $f(p^2)\sim m^2 e^{p^2/\Lambda^2}$, the cosmological constant scales logarithmically with cutoff, $\lambda(\epsilon)\sim \Lambda^4\ln^3(m\epsilon)$. This behavior is radically different from the standard UV behavior of cosmological constant.
	\item 	Although the degree of UV divergence becomes milder when $f(p^2)$ is switched on the divergence never disappears completely. In fact all terms in the effective action that were UV divergent in a theory with propagator $1/p^2$ remain divergent if propagator is perturbed by $f(p^2)$.
	\item  All what we have just said is true for entanglement entropy. The entropy,as we have found,  is  proportional to the area in the same way as in a theory with standard propagator although its dependence on the UV cutoff modifies respectively. However, the entropy never becomes UV finite even if the modification in the propagator grows exponentially with $p^2$.  In fact, as we have observed,  the UV divergence of the entropy is the same as that of Newton's constant.
	\item In the case of theories with broken Lorentz symmetry there is a set of couplings in the effective gravitational action corresponding to the geometric terms which are determined either  by the space-like metric or by the extrinsic curvature. We have obtained the dependence of the couplings on the UV cutoff. We show that under  certain  mild restrictions on the dimension of the Lifshitz operator which breaks the Lorentz	 symmetry the couplings of the extrinsic curvature in the action become UV finite. This is the first and only example we know when the couplings which were UV divergent in the theory with standard propagator become UV finite  when the propagator is improved for large momentum.
	 \item Entanglement entropy in a generic theory with broken Lorentz symmetry described by  propagator (\ref{i4}) has been calculated and we have found that, quite surprisingly, it depends on the UV cutoff $\epsilon$ in the same way as in Lorentz invariant theory with propagator (\ref{i1}) provided function $F$ is the same in both cases. The area term in the entropy thus does not seem to be sensitive to the nature, Lorentz invariant or not, of the UV modification of the theory. Also we note that propagators of the type (\ref{i4}) are typical for condensed matter systems. We anticipate that our results of Section 6.2 may have important applications in studying entanglement entropy in these systems.
\end{itemize}

Our findings may have interesting phenomenological applications which we plan to discuss elsewhere.

\bigskip
\section*{Acknowledgements}
We thank O. Lysovyi, S. Nicolis, M. Volkov and other colleagues in Tours for useful discussions.
The postdoctoral position of D.N. is financed by the University of Tours. S.S. is grateful to the Theory Division at CERN for the hospitality extended to him while this work was in progress. D.N. was also supported by the RFBR grant 08-02-00725.

\newpage
\appendix
\section{Heat kernel of operator $F(\Box_g)$ in momentum representation}\setcounter{equation}0
In this appendix we perform perturbative calculation of the curvature term in heat kernel. First we decompose the $d$-metric
 \[ g^{\alpha\beta}(X)=\delta^{\alpha\beta}+h^{\alpha\beta}(X)\,, \]
where $\delta^{\alpha\beta}$ is metric on flat spacetime and $h^{\alpha\beta}$ is a small perturbation. Then  the covariant operator
 \[ \Box_g=-{1\o \sqrt{g}}\partial_\alpha (\sqrt{g}g^{\alpha\beta}\partial_\beta) \]
is presented as a sum of the Laplace operator in flat space and a perturbation
 \begin{eqnarray}
 &&\Box_g=\Box_0+\delta \Box~,~~\Box_0=-\partial_\alpha^2~,
 \nonumber \\
 &&\delta\Box=-\partial_\alpha(h^{\alpha\beta}(X)\partial_\beta )+{1\o 2}(\partial^\alpha h(X))\partial_\alpha~, \lb{4.1}
 \end{eqnarray}
where $h(X)=\delta_{\alpha\beta}h^{\alpha\beta}(X)$. In what follows it is useful to apply a Fourier decomposition for the perturbation
 \begin{equation}
  h^{\alpha\beta}(X)={1\o (2\pi)^d}\int d^d k\, e^{-ikX}h^{\alpha\beta}(k) \,, \lb{4.2}
 \end{equation}
where $kX=k_\mu X^\mu$. We will assume that the function $F(\Box)$ can be presented as a
Taylor series
 \begin{equation} F(\Box_g)=\sum_{n=0}^\infty {F^{(n)}(0)\o n!}\Box_g^n~. \lb{4.3}  \end{equation}
A variation of $n$-th power of the Laplace operator is
 \begin{equation} \delta \Box_g^n=\sum_{j=0}^{n-1}\Box_0^{n-j-1}\delta\Box\, \Box_0^j~. \lb{4.4} \end{equation}
Respectively we have for the variation of $F(\Box_g)$,
 \begin{equation} \lb{FF}
  \delta F(\Box_g)=\sum_{n=0}^\infty {F^{(n)}(0)\o n!}\sum_{j=0}^{n-1}\Box_0^{n-j-1}\delta\Box\, \Box_0^j \,.
 \end{equation}
To first order in perturbation we can represent the heat kernel
in the form
 \begin{equation} \lb{4.5}
  K(s,X,X')=K_0(s,X,X')+K_1(s,X,X') \,,
 \end{equation}
where
 \begin{equation} \lb{4.6}
  K_0(s,X,X')=\int {d^dp\o (2\pi)^d}e^{-ip(X-X')}\,e^{-sF(p^2)}
 \end{equation}
is the heat kernel in flat space, $(\partial_s+\Box_0)K_0(s,X,X')=0$, and the term
 \begin{equation} \lb{4.7}
  K_1(s,X,X')=\int {d^dp\o (2\pi)^d}e^{-ip(X-X')}K_1(s,p,X'),
 \end{equation}
is  due to the perturbation. The heat equation to first order in perturbation reduces to equation
 \begin{equation} \lb{4.41}
  (\partial_s+F(\Box_0))K_1+\delta F(\Box_g)K_0=0
 \end{equation}
which can be easily solved to give the function $K_1(s,X,X')$ in the form (\ref{4.7}) with
 \begin{eqnarray} \label{4.8}
 && K_1(s,p,X')
  =\int {d^dk\o (2\pi)^d}e^{-ikX'}D(p,k){\cal F}(p,k)H(p,k)~,
 \\
 && D(p,k) ={e^{-sF(p^2)}-e^{-sF((p-k)^2)}\o F(p^2)-F((p-k)^2)}~,
 \nonumber\\
 && {\cal F}(p,k) =\sum_{n=0}^\infty {F^{(n)}\o n!}\sum_{j=0}^{n-1}(p^2)^{n-j-1}((p-k)^2)^j~,
 \nonumber \\
 && H(p,k)
  =(p_\alpha(p-k)_\beta-{1\o 2}\delta_{\alpha\beta}k^\sigma(p-k)_\sigma)h^{\alpha\beta}(k)~.
\nonumber
 \end{eqnarray}
The sum over $j$ in expression for function ${\cal F}(p,k)$ can be easily evaluated
 \[ \sum_{j=0}^{n-1}(p^2)^{2(n-j-1)}((p-k)^2)^j={((p-k)^2)^n-(p^2)^n\o (p-k)^2-p^2} \,. \]
Noting  the Taylor series
 \begin{eqnarray}
 &&\sum_{n=0}^\infty{F^{(n)}(0)\o n!}((p-k)^2)^n=F((p-k)^2)~,\nonumber\\
 &&\sum_{n=0}^\infty{F^{(n)}(0)\o n!}(p^2)^n=F(p^2)~. \nonumber
 \end{eqnarray}
we obtain a simple expression
 \begin{equation} \lb{4.9}
  {\cal F}(p,k)={F((p-k)^2)-F(p^2)\o (p-k)^2-p^2}
 \end{equation}
so that the product of functions $D(p,k)$ and ${\cal F}(p,k)$ in (\ref{4.8}) becomes
 \begin{equation} \lb{4.10}
  D(p,k){\cal F}(p,k)={e^{-sF(p^2)}-e^{-sF((p-k)^2)}\o p^2-(p-k)^2}~.
 \end{equation}

The expression (\ref{4.8}) with (\ref{4.10}) is linear in perturbation $h^{\alpha\beta}$. On the other hand, it is
nonperturbative in momentum $k^\alpha$ or, in other words, in derivative of $h^{\alpha\beta}(X)$. Using (\ref{4.8}) we can in principle obtain the complete result for   the part of the effective action that is linear in the perturbation. Written in a covariant form this would give us all terms in the effective action which are linear in the Riemann tensor and containing arbitrary number of derivatives.

Since we present the effective action in a form of decomposition in number of derivatives of metric the expression for the heat kernel $K_1(s,X,X')$ (\ref{4.8}) should be decomposed in a Taylor series in momentum $k^\alpha$. First we note that
 \begin{eqnarray} \lb{4.11}
  &&{e^{-sF(p^2)}-e^{-sF((p-k)^2)}\o p^2-(p-k)^2}=\partial_{p^2}[e^{-sF(p^2)}]
 \\
  &&\qquad\qquad +\,{1\o 2}\partial^2_{p^2}[e^{-sF(p^2)}]((p-k)^2-p^2)  +{1\o 6}\partial^3_{p^2}[e^{-sF(p^2)}]((p-k)^2-p^2)^2+..\;.
  \nonumber
 \end{eqnarray}
With this formula we find  a decomposition in $k^\alpha$
 \begin{equation} \lb{4.12}
  (D\cdot {\cal F}\cdot H)(p,k)=A^{(0)}(p)+A^{(1)}_\alpha(p) k^\alpha+A^{(2)}_{\alpha\beta}(p)k^\alpha k^\beta+..\;,
 \end{equation}
where
 \begin{eqnarray} \lb{4.13}
  &&A^{(0)}(p)=\partial_{p^2} [e^{-sF(p^2)}]\, p_\alpha p_\beta \, h^{\alpha\beta}(k),
 \\ \lb{4.14}
  &&A^{(1)}_\alpha(p)=\partial_{p^2} [e^{-sF(p^2)}](-p_\beta h^\beta_{\, \alpha}(k) - {1\o 2}h(k)p_\alpha) - \partial_{p^2}^2[e^{-sF(p^2)}]\, p_\alpha p_\beta p_\gamma \, h^{\beta\gamma}(k),
 \\ \label{4.15}
  &&A^{(2)}_{\alpha\beta}(p)
   =\!\!{1\o 2}\partial_{p^2}^2[e^{-sF(p^2)}](\delta_{\alpha\beta}p_\gamma p_\sigma h^{\gamma\sigma}(k)
   \!\!\!+\, h(k)p_\alpha p_\beta+2p_\alpha p_\sigma h^{\sigma}_{\, \beta}(k))
 \\
  &&\qquad\qquad+ \;{2\o 3}\partial^3_{p^2}[e^{-sF(p^2)}]p_\alpha p_\beta p_\gamma p_\sigma h^{\gamma\sigma}(k).
   \nonumber
 \end{eqnarray}
Now we are in a position to calculate the trace
 \begin{equation}
  \Tr K_1 (s) =\int d^dX \,K_{1}(s,X,X'{=}X) =\int d^dX\int{d^dp\o (2\pi )^d}\, K_{1}(s,p,X) \,.
 \nonumber
 \end{equation}
Using the formulae for the integration over $p$ which we list in Appendix B and returning to the coordinate representation we find
 \begin{eqnarray}
  \int\!\! {d^dp\o  (2\pi)^d}K_1(s,p,X) = \frac1{(4\pi)^{d/2}} \Big(\! {-}P_{d}(s)\frac12 h(X)
   + P_{d-2}(s)\frac16 \big({-}\partial_\za \partial_\zb h^{\alpha\beta}(X){+} \partial^\za \partial_\za h(X)\big)\!\Big).
 \label{4.16}
 \end{eqnarray}
In terms of perturbation over flat metric the Ricci scalar reads
  \[ R={-}\partial_\za\partial_\zb h^{\za\zb}(X){+}\partial^\za \partial_\za h(X) \,.\]
Thus the last term in (\ref{4.16}) is proportional to the Ricci scalar. Taking that in this approximation in $h^{\alpha\beta}$ we have that $\sqrt{g}=1-{1\o 2}h(X)+O(h^2)$ and $R=\sqrt{g}R+O(h^2)$ we finally obtain (\ref{4.17})
 \begin{eqnarray}
  \Tr K(s)=\Tr K_0(s)+\Tr K_1(s)
   = \frac1{(4\pi)^{d/2}} \Big(P_{d}(s) \int\! d^dX\sqrt{g}+
   P_{d-2}(s)\int\! d^dX\sqrt{g}\,\frac16 R  + ..\Big)
 \nonumber \label{4.17-A}
 \end{eqnarray}
for the trace of the heat kernel of operator $F(\Box_g)$.

\bigskip

\section{Some Integrals in Momentum Space}\setcounter{equation}0

For $d>2$ the following is true
 \begin{eqnarray} \lb{a1}
  \frac1{(2\pi)^d} \int d^dp\, \partial_{p^2}e^{-sF(p^2)}
   &&\!\!\!\!\!\!\!= - \frac1{(4\pi)^{d/2}} P_{d-2}(s) ,
 \\ \lb{a2}
  \frac1{(2\pi)^d} \int d^dp\, \partial_{p^2}[e^{-sF(p^2)}]p_\alpha p_\beta
   &&\!\!\!\!\!\!\!= - \frac1{(4\pi)^{d/2}}  \frac{\delta_{\alpha\beta}}{2} P_{d}(s) ,
 \\ \lb{a3}
  \frac1{(2\pi)^d} \int d^dp\, \partial^2_{p^2}[e^{-sF(p^2)}]p_\alpha p_\beta
   &&\!\!\!\!\!\!\!= \;\frac1{(4\pi)^{d/2}}  \frac{\delta_{\alpha\beta}}{2} P_{d-2}(s),
 \\ \lb{a4}
  \frac1{(2\pi)^d} \int d^dp\, \partial^3_{p^2}[e^{-sF(p^2)}]p_\alpha p_\beta p_\sigma p_\gamma
   &&\!\!\!\!\!\!\!= -\frac1{(4\pi)^{d/2}} \frac{(\delta_{\alpha\beta}\delta_{\gamma\sigma} {+}\delta_{\alpha\gamma}\delta_{\beta\sigma} {+}\delta_{\alpha\sigma}\delta_{\gamma\beta})}{4}  P_{d-2}(s) ,
 \\ \lb{a5}
  \frac1{(2\pi)^d} \int d^dp\; \partial^n_{p^2}[e^{-sF(p^2)}]p_{\alpha_1} p_{\alpha_2}
     .. p_{\alpha_{2m+1}}
   &&\!\!\!\!\!\!\!=0 \,,
  \qquad n,m \in \mathbb{Z}\;,
 \end{eqnarray}
where (\ref{4}) $P_{n}(s)=\frac2{\GammaF{\frac{n}2}} \int_0^\infty dp\, p^{n-1}\, e^{-sF(p^2)}$.

\bigskip

\section{Heat kernel of operator $\Delta_\tau+F(\Delta_x)$}\setcounter{equation}0
{
 \renewcommand{\apt}{\sigma}       
 \renewcommand{\Nl}{N}               
 \renewcommand{\pBt}{{\Delta_\tau}} 

In this appendix we want to comment on explicit calculation of the
effective action on a ``curved spacetime'', in  this case the
lapse function $N$ and the $(d{-}1)$-metric $g_{ij}$ in (\ref{5.1})
are arbitrary functions of $\tau$ and $x^i,\; i=1,2,..,d{-}1$. For the  heat kernel of operator $\Delta_\tau{+}F(\Delta_x)$
we want to  use a representation similar to (\ref{E1}). Then by introducing the constantly rescaled lapse function $\tilde{N}^2(\tau,x)={\sigma\o s}\,N^2(\tau,x)$  the heat kernel
$e^{-(s\Delta_\tau+\sigma\Delta_x)}$ turns into the heat kernel
$e^{-\sigma\tilde{\Box}}$ of \emph{covariant} $d$-dimensional Laplace
operator\footnote{
 We note that in contrast to the case of a cone space for smooth manifold case the heat kernel of Laplacian  with complex metric is defined straightforwardly.  And all further calculations are also justified since the underlying DeWitt's procedure of obtaining heat kernel expansion does not rely on realness of metric coefficients.} $\tilde{\Box}=\tilde{\Delta}_\tau+\Delta_x$ built from a new metric $\tilde{G}_{\zm\zn} =\{{\tilde N}^2, g_{ij}\}$. The advantage of this representation is that we can use the well-known
Schwinger-DeWitt expansion of the heat kernel of the Laplace operator.

The formula (\ref{E1}) however cannot be directly
applied since the operators $\Delta_\tau$ and $\Delta_x$ do not
commute if the lapse function $N$ and the metric $g_{ij}$ are
nontrivial functions.
 In this case one has to decouple exponents of these operators, i.e. represent  $e^{-s(\Delta_\tau+F(\Delta_x))}$
as a product of two exponents $e^{-sF(\Delta_x)}$  and
$e^{-s\Delta_\tau}$, then use the representation (\ref{E1}) for
 $e^{-sF(\Delta_x)}$ and finally combine two exponents
$e^{-s\Delta_\tau}$ and $e^{-\sigma\Delta_x}$ into one
$e^{-(s\Delta_\tau+\sigma\Delta_x)}$.
The noncommutativity of $\Delta_\tau$ and $\Delta_x$ at each step leads to pre-factors which are local differential operators and are constructed from commutators of the operators. Here one can use a Zassenhaus formula.The Zassenhaus formula is analogue of Baker-Campbell-Hausdorff formula which arranges the result in a different way suitable for our needs
 \[ e^{s(A+B)}=e^{s A}e^{s B}e^{s^2 C_2}..e^{s^n C_n}.. \,, \]
where $A,B$ -- some operators and $C_n$ are linear combinations of $n$ commutators of  operators $A$ and $B$, for instance one has \cite{ZassenhausFormula}
 \[ C_2=-\frac12 [ A,B]\,,\;C_3= \frac16 (2[B,[A,B]]+[A,[A,B]])\,. \]

In our case we use both direct and transposed variants of the Zassenhaus formula and expand exponents with commutators  (with $C_2$, $C_3$,.. ) in series. This allows us to effectively decouple and recouple exponents of noncommuting operators in question. Here we choose to place prefactors to the left of heat kernels assuming all derivatives act to the right
  \begin{equation}
   e^{-\spt(\Delta_\tau+F(\Delta_x))}
   = \int_0^\infty dq\; e^{-sF(q)} \int_{\!{-}i\infty}^{i\infty} \frac{d \apt}{2\pi i}\; e^{\apt q}\;
   \hat{Q}\; e^{-(\spt\Delta_\tau+\apt\Delta_x)}~,
  \label{H}
  \end{equation}
where
  \begin{eqnarray}
   & \hat{Q}(\spt,\apt,q)
    &= 1 \; - \; [\pBx,\pBt] \Big( \frac12\spt^2 F'(q)-  \frac12\spt\apt \Big) +\, [\pBt,[\pBx,\pBt]]\,\Big( \frac16 \spt^3 F'(q) - \frac16 \spt^2\apt  \Big)
   \nonumber\\
   &&\quad +\, [\pBx,\pBt] [\pBx,\pBt] \Big(\, \frac18\spt^4 \,F'(q)F'(q) + \frac18\spt^2\apt^2 -\frac14\spt^3\apt F'(q) -\frac16 \spt^3 F''(q)\,\Big)
    \nonumber\\
   &&\quad +\, [\pBx,[\pBx,\pBt]]\Big(-\frac14\spt^2 F''(q) +\frac13\spt^3 F'(q)F'(q) - \frac13\spt\apt^2 \Big)
     \;  + O([\,,]^3)\, .
   \label{Q_initial}
  \end{eqnarray}
We note that our aim is to present trace of the heat kernel (\ref{H}) as an expansion in number of derivatives acting on the metric components, i.e. on functions $N(\tau,x)$ and $g_{ij}(\tau,x)$. Commutators in (\ref{Q_initial}) are differential operators whose coefficients are (products of) derivatives of the metric. Counting number of derivatives in these coefficients one observes that the commutator $[\Delta_x,\Delta_\tau]$ is proportional to at least one derivative which acts on metric components. Each additional commutation with either $\Delta_x$ or $\Delta_\tau$ increases the number of such derivatives. In this context $O([\,,]^3)$ produce terms which contain at least three derivatives. Such counting is closely related to the notion of \emph{background dimensionality} \cite{Barvinsky:2009cb, Barvinsky:1985an}.

When being integrated (\ref{H}) not all terms in (\ref{Q_initial}) are independent which makes such representation of $Q(\spt,\apt,q)$ \emph{nonunique}. To reduce to unique representation we use integration by parts over $q$ to remove all explicit dependence on $\apt$ which enters (\ref{Q_initial}) in positive powers\footnote{Positive powers of $\apt$ in (\ref{Q_initial}) can be represented as powers of $\partial_q$ acting on $e^{q\apt}$ which after integration by parts effectively act on $e^{-s F(q)}$ leading to powers of $\spt$ and derivatives of $F(q)$. Note also that no terms at $q=0$ arise while integrating by parts that can be seen from the precise representation of (\ref{H}) in terms of contour integrals. In context of the formal derivation given in Section 5, these boundary terms (after taking the trace) will be proportional to integrals over $\apt$ of sufficient negative powers of the latter. These integrals vanish since they start and end at infinities and does not pass over zero or a cut in complex $\apt$-plain.}. The reduced form  of operator $\hat Q$ then reads
 \begin{equation} \label{Q_final}
  \hat{Q}(\spt,q)
   = 1 - \frac1{24} [\pBx,\pBt ][\pBx,\pBt ] s^3 F''(q)
     + \frac1{12} [\pBx,[\pBx,\pBt]] s^2 F''(q)
   +O([\,,]^3) .
 \end{equation}
In particular, in the reduced representation (\ref{Q_final}) all terms containing  $F'(q)$ disappear. This is just what one expects since in the Lorentz invariant  case (when $F'(q)=1$, $F''(q)=F'''(q)=..=0$) the operator  $\hat{Q}\equiv 1$ as can be seen from (\ref{H}).
\\

Commutators of operators $\pBt$ and $\pBx$ are calculated straightforwardly:
  \begin{eqnarray}    \label{Commutators}
     &[\pBx,\pBt][\pBx,\pBt]
      &=  16 \frac1{\Nl^2} k^{{i}{j}} k^{{l}{m}} \nabla_{i}\nabla_{j}\nabla_{l}\nabla_{m}\partial_\tau^2
        - 32 \frac1{\Nl^3} k^{{i}{j}}\nabla^{l}\!\ln\Nl\, \nabla_{i}\nabla_{j}\nabla_{l} \partial_\tau^3
        \\
     &&\quad  + 16 \frac1{\Nl^4} \nabla^{i}\!\ln\Nl \nabla^{j}\!\ln\Nl\, \nabla_{i}\nabla_{j}\partial_\tau^4
        +    O(\CMdim{3}),
        \nonumber\\ 
     &[\pBx,[\pBx,\pBt]]
      &= - 8 k^{{i}{j}}k^{{l}{m}} \nabla_{i}\nabla_{j}\nabla_{l}\nabla_{m}
         - 8 \frac1{\Nl} \nabla^{{m}} k^{{i}{j}} \; \nabla_{i}\nabla_{j}\nabla_{m} \partial_\tau
         + 24 \frac1{\Nl} k^{{i}{j}} \nabla^{{m}}\!\ln\Nl \; \nabla_{i}\nabla_{j}\nabla_{m}\partial_\tau
        \nonumber\\
        &&\quad + 8 \frac1{\Nl^2}\nabla^{{m}} \nabla^{{i}}\!\ln\Nl \; \nabla_{i}\nabla_{m} \partial_\tau^2
         - 16 \frac1{\Nl^2}\nabla^{{m}}\!\ln\Nl \nabla^{{i}}\!\ln\Nl \; \nabla_{i}\nabla_{m} \partial_\tau^2
         +O(\CMdim{3}).
   \nonumber
   \end{eqnarray}
In expressions above $k_{{i}{j}}$ is the extrinsic curvature of constant time hypersurfaces (\ref{5.6}), $\nabla$ is space-like covariant derivative and $O(\CMdim{3})$ denotes terms -- differential operators -- which in its coefficients contain at least $3$ derivatives acting on the $d$-dimensional metric. These terms do not affect the result for heat kernel trace which we restrict here to $O(\CMdim{2})$ order.

As we have already noticed the operator
$e^{-(s\Delta_\tau+\sigma\Delta_x)}$ can be viewed as the heat
kernel $e^{-\sigma\tilde{\Box}}$ of Laplace operator
$\tilde{\Box}$ for the new metric
$\tilde{G}_{\mu\nu}=\{\tilde{N}^2,g_{ij}\}$ with the rescaled
lapse function $\tilde{N}^2(\tau,x)={\sigma\o s}N^2(\tau,x)$. The
small $\sigma$ expansion of this heat kernel is well-known \cite{DeWitt} and can be
used in our calculation,
 \begin{equation}
  e^{-\sigma\tilde{\Box}} \delta(\tau,x;\tau',x')
   = {\tilde{D}^{1/2}(\tau,x;\tau',x')\o (4\pi\sigma)^{d/2}}
    e^{-{\tilde{\pmb\sigma}(\tau,x;\tau',x')\o 2\sigma}}
    \sum_{n=0}^\infty\sigma^n \tilde{a}_n(\tau,x;\tau',x') ~,
 \label{HKansatz}
 \end{equation}
where $\tilde{\pmb\sigma}(\tau,x;\tau',x')$ and $\tilde{D}(\tau,x;\tau',x')$ are  world function and van Vleck-Morette determinant (computed for metric $\tilde{G}_{\mu\nu}$) respectively  \cite{DeWitt}. The first two Schwinger-DeWitt coefficients $\tilde{a}_n$ at coinciding arguments read
 \begin{equation}
  \tilde{a}_0(\tau,x;\tau,x)=1~,\;\;
  \tilde{a}_1(\tau,x;\tau,x)={1\o 6}R(\tilde{G})
   = \frac16\big(R^{d{-}1}(g)+\tilde{k}^2-\tilde{k}_{ij}\tilde{k}^{ij}+ {\it t.d.}\big)\,,
 \label{D5}
 \end{equation}
where $t.d.$ stands for total derivative terms which we neglect hereafter.

The next  step is to calculate the trace of $\hat Q$ (\ref{Q_final}), (\ref{Commutators}) acting on  expansion (\ref{HKansatz}). Restriction to terms which contain not more than two derivatives of the metric considerably simplifies the calculation.

The first term in $\hat Q$ (\ref{Q_final}) acting on (\ref{HKansatz}) gives
 \begin{eqnarray}
  \frac{1}{(4\pi\sigma)^{d/2}}\int d\tilde{\mu} \left[ 1\;+\; \apt \frac16\big(\,R^{(d{-}1)}(g) + {\tilde k}^2 - \tilde{k}_{ij}\tilde{k}^{ij}\big) \right]\,,
 \label{*D}
 \end{eqnarray}
where $d \tilde{\mu}=d\tau d^{d{-}1}\!x \,\tilde{N}\sqrt{g}$.

The rest two terms in $\hat Q$ (\ref{Q_final})  are already  $O(\CMdim{2})$. Thus we have to look at only at terms which when acting on (\ref{HKansatz}) do not produce extra derivatives of metric.  This happens only when $a_0$ coefficient is left in  (\ref{HKansatz}) and when ``free'' derivatives of (\ref{Commutators}) act by pairs on the world function $\tilde{\pmb\sigma}$ in  (\ref{HKansatz}) thus producing the components of metric  but not its derivatives. One needs to know only the following structures at coinciding arguments: $\nabla_\za\nabla_\zb \tilde{\pmb\sigma}=g_{\za\zb}$, $\nabla_\za\partial_\tau \tilde{\pmb\sigma}=0$, $ \partial_\tau \partial_\tau \tilde{\pmb\sigma}=\tilde{\Nl}^2$. Up to total derivatives this gives
    \begin{eqnarray}
      &\frac{1}{(4\pi\sigma)^{d/2}}
      \int d\tilde{\mu} &\left[ \frac1{12} \frac{\spt^3}{\apt^3} F''(q) \frac{\tilde{\Nl}^2}{\Nl^2}\big( k^2 + 2 {k}_{ij}{k}^{ij} \big)
        +\frac14 \frac{\spt^3}{\apt^3} F''(q) \frac{\tilde{\Nl}^4}{\Nl^4} \frac{\nabla \Nl \nabla \Nl}{\Nl^2}\right.
       \nonumber\\
       &&\left.- \frac16 \frac{\spt^2}{\apt^2} F''(q)  \big( {k}^2 + 2 {k}_{ij}{k}^{ij} \big)
       -  \frac12\frac{\spt^2}{\apt^2} F''(q) \frac{\tilde{\Nl}^2}{\Nl^2} \frac{\nabla \Nl \nabla \Nl}{\Nl^2}\right] \,.
    \label{*D1}
    \end{eqnarray}

Returning back to original metric (i.e. replacing $\tilde{N} {=} \sqrt{\apt/\spt}\, N$ and $\tilde{ k}_{ij} {=} \sqrt{\spt/\apt}\, k_{ij}$) and combining together (\ref{*D}) and (\ref{*D1}) one finally comes to\footnote{
We independently confirmed this result by performing a perturbative calculation using momentum representation (similar to that of Appendix A but involving second order in perturbation).
}
    \begin{eqnarray}
      &\Tr \,\hat{Q}\, e^{-\apt\tilde{\Box}}
      = & \displaystyle{\frac{1}{(4\pi)^{d/2}\apt^{(d{-}1)/2}\spt^{1/2}}}
      \int d{\mu} \;\Big[1+  \frac16 \apt R^{d{-}1}(g)+ \frac16 \spt \big({k}^2 - {k}_{ij}{k}^{ij}\big)
      \nonumber\\
      &&  \qquad\qquad - \frac1{12} \frac{\spt^2}{\apt^2} F''(q) \big( {k}^2 + 2{k}_{ij}{k}^{ij} \big)   - \frac14 \frac{\spt}{\apt} F''(q) \frac{\nabla \Nl \nabla \Nl}{\Nl^2}  + O(\CMdim{3})\Big]\, ,
    \nonumber
    \end{eqnarray}
where $d {\mu}=d\tau d^{d{-}1}\!x \,{N}\sqrt{g}$.

Integration  over auxiliary proper time $\apt$ (which gives just certain powers of $q$) and subsequent integration over $q$ completes calculation of (\ref{7.1}), (\ref{7.2}).


\end{document}